\documentclass[preprint,pre,aps,amsmath,amssymb,a4paper,superscriptaddress,nofootinbib,11pt,tightenlines]{revtex4-2} 

\usepackage{graphicx}	
\usepackage{dcolumn}	
\usepackage{bm}			
\usepackage{xcolor}		
\usepackage{hyperref}	
\usepackage[capitalize]{cleveref}	
\usepackage{amsthm} 	
\usepackage{lineno} 	

\usepackage{url}							

\newcommand{\angles}[1]{\langle{#1}\rangle}

\newcommand{\ignore}[1]{}					

\newtheorem{theorem}{\noindent Theorem}

\hypersetup{				
	colorlinks   = true, 	
	urlcolor     = blue, 	
	linkcolor    = blue, 	
	citecolor    = blue     
}
\crefformat{equation}{Eq.~#2(#1)#3}				
\crefformat{figure}{Fig.~#2#1#3}				
\crefformat{section}{Sec.~#2#1#3}				

\begin{document}
\title{Lorentzian geometry and variability reduction in airplane boarding: Slow passengers first outperforms random boarding}
\author{Sveinung Erland}
\affiliation{Department of Maritime Studies,\\ {Western Norway University of Applied Sciences, N-5528 Haugesund, Norway}}
\email[Author to whom correspondence should be addressed: ]{sver@hvl.no}
\author{Jevgenijs Kaupu\ifmmode \check{z}\else \v{z}\fi{}s}
\affiliation{Faculty of Materials Science and
	Applied Chemistry, Institute of Technical Physics, Riga Technical University, LV-1048 Riga, Latvia}
\affiliation{Institute of Mathematical Sciences and Information Technologies, University of Liepaja, LV-3401 Liepaja, Latvia}
\author{Albert Steiner}
\affiliation{Institute of Data Analysis and Process Design, Zurich University of Applied Sciences ZHAW, 8401 Winterthur, Switzerland}
\author{Eitan Bachmat}
\affiliation{Department of Computer Science, Ben-Gurion University, Beer-Sheva 84105, Israel}
\date{\today}

\begin{abstract}
Airlines use different boarding policies to organize the queue of passengers waiting to enter the airplane. We analyze three policies in the many-passenger limit by a geometric representation of the queue position and row designation of each passenger, and apply a Lorentzian metric to calculate the total boarding time. The boarding time is governed by the time each passenger needs to clear the aisle, and the added time is determined by the aisle-clearing time distribution through an effective aisle-clearing time parameter. The non-organized queues under the common random boarding policy are characterized by large effective aisle-clearing time. We show that, subject to a mathematical assumption which we have verified by extensive numerical computations in all realistic cases, the average total boarding time is always reduced when slow passengers are separated from faster passengers and the slow group is allowed to enter the airplane first. 
This is a universal result that holds for any combination of the three main governing parameters: the ratio between effective aisle-clearing times of the fast and the slow group, the fraction of slow passenger, and the congestion of passengers in the aisle.
Separation into groups based on aisle-clearing time allows for more synchronized seating, but the result is non-trivial, as the similar fast-first policy --- where the two groups enter the airplane in reverse order --- is inferior to random boarding for a range of parameter settings. The asymptotic results conform well with discrete-event simulations with realistic number of passengers, and both the slow-first and the fast-first policies have the ability to perform unboundedly better than random boarding. Parameters based on empirical data, with hand luggage as criteria for separating passengers into the slow group, give a 13\% reduction in total boarding time for slow first compared to random boarding.
\end{abstract}


\maketitle
\clearpage
\section{Introduction}\label{sec:introduction}
Determination and optimization of the macroscopic properties of complex systems are of importance in many fields. In airplane boarding the main observable is the boarding time, which is the time it takes to get all passengers seated. The boarding time is affected by several factors including sequencing of the queue, passenger interactions and the time each passenger needs to clear the aisle. Moreover, the dynamics of the passenger queue are becoming increasingly complex as the number of passengers $N$ increases.

While the quantification of properties of complex systems often requires extensive simulations, the airplane boarding process can be analyzed in terms of a geometrical representation of the queue position and the row designation of the passengers. When the number of passengers $N\rightarrow\infty$, a flat Lorentzian metric enables the boarding time to be expressed in analytical terms, which also enables rough estimates for finite number of passengers \cite{Bachmat/Berend/Sapir/Skiena/Stolyarov:2006,Bachmat:2014}. The boarding time is found to be of leading order $\sqrt{N}$, and it is scaled by a pre-factor that is governed by three main parameters (that will be explained later). The analytical expressions enable a direct optimization over these parameters \cite{Frette/Hemmer:2012,Bernstein:2012,Martins/Kaupuzs/Mahnke:2013,Baek/Ha/Jeong:2013,Bachmat/Khachaturov/Kuperman:2013,Mahnke/Kaupuzs/Brics:2015,Bachmat:2019, Erland/Kaupuzs/Frette/Pugatch/Bachmat:2019}.

In this paper we set particular focus on the effect of varying aisle-clearing times. When the aisle-clearing time $X$ of each passenger is stochastic, there exists a constant $\tau_X$ such that the asymptotic boarding time $T$ is the same as if the aisle-clearing time was constant for all passengers with value $X=\tau_X$. We therefore call $\tau_X$ the effective aisle-clearing time. The parameter is another macroscopic property of airplane boarding, and it is similar to what in material science is described as effective transport and optical properties for composite materials \cite{Stroud:1975,Fan:1996,Wang/Pan:2008,Braun/Pilon:2006}. As in material sciences, the computation of $\tau_X$ is not trivial, and one of the main contributions of this work is to devise a method for precise estimation of the parameter. We utilize the property that $\tau_X$ depends on the aisle-clearing time distribution only, and interestingly, this also reduces the task to the mathematical problem of finding the heaviest increasing sub-sequence in a permutation \cite{Jacobson/Vo:1992}.

We study the airplane boarding process from the passengers have lined up in a queue outside the airplane until the last {passengers are} seated at their reserved seats. Empirical data shows that a reduction of the boarding time would reduce the total turnaround time and hence reduce airline costs \cite{Neumann:2019}. Still, most passengers have experienced that boarding is a seemingly chaotic process where most of the time is spent in the queue waiting. 

The way airlines organize the queue prior to boarding is called a boarding policy. The random boarding policy, with its completely unorganized queues, is surprisingly common, in particular for intra-continental flights in Europe \cite{Delcea/Cotfas/Salari/Milne:2018}. There have been several efforts to find optimal policies \cite{Jaehn/Neumann:2015}, but airlines hesitate to apply the solutions since optimal queues require that individual passengers adhere to specific positions in the queue \cite{vanLandeghem/Beuselinck:2002,Steffen:2008,Steffen:2008b,Steffen/Hotchkiss:2012}, and such detailed regulation could be detrimental to passenger satisfaction. Aspects such as priority boarding of passengers with high revenue tickets will also limit the potential for optimization.

Many airlines do enforce boarding policies with weaker restrictions, where passengers typically are divided into two or more groups. A common policy is the back-to-front policy where the first group to enter the queue is passengers with designated seats in the back of the airplane. The next group will be seated at consecutive rows closer to the front, and so on. In spite of these efforts, the policy tends to increase boarding time compared to random boarding \cite{Bachmat/Khachaturov/Kuperman:2013}. A less common policy is the window-middle-aisle policy where groups designated to window seats are asked to sit first, followed by the middle-seats and so on. However, here passengers that travel together run the risk of getting separated during boarding.

Other applied policies involve priority boarding where, e.g., groups consisting of small children or other needing assistance are asked to enter first. This is actually a variant of the slow-first policy where passengers who are expected to use a long time to clear the aisle (the slow group) are allowed to enter first, followed by the faster passengers. The universal result in \cite{Erland/Kaupuzs/Frette/Pugatch/Bachmat:2019} states that the slow-first policy is always superior to the opposite fast-first policy. In this paper we prove another universal result, namely that slow first is also universally better than the random boarding policy. The result is a confirmation of the simulation-based result in \cite{Audenaert/Verbeeck/Berghe:2009}. We quantify the result by taking parameters from empirical data where the slow group is defined to be passengers with overhead bin luggage. Other boarding policy models that take the aisle-clearing time into account exist, and even though they have the potential to reduce the boarding time even further, they all assign specific position in the queue for each passenger \cite{Qiang/Jia/Xie/Gao:2014, Milne/Kelly:2014,Milne/Salari:2016,Notomistaetal:2016}. And, as noted above, this is most likely not beneficial for the customer satisfaction.

The structure of the paper is as follows. 
In \cref{sec:boardingprocess} we describe the boarding process, followed by a summary of the main results in \cref{sec:mainresults}. 
Main parameters of the boarding process, its geometric representation and the blocking chains that lead to 
the asymptotic boarding time in the many-passenger limit ($N\rightarrow \infty$), are presented in \cref{sec:model}.
In \cref{sec:CX} we present bounds for the effective aisle-clearing time $\tau_X$, that is needed for computing the asymptotic boarding time when the aisle-clearing time $X$ varies between passengers within each group. $\tau_X$ is particularly important for random boarding, and we prescribe an algorithm that enables precise estimation of $\tau_X$ and argue that $\sqrt{\langle X^2 \rangle}$ serves as a lower bound for $\tau_X$.
In \cref{sec:RAvsSF} we present and compare analytical results for the asymptotic boarding time of the slow-first (SF), the fast-first (FF) and the random boarding (RA) policies. We also demonstrate by simulations that the large-$N$ limit results hold for a realistic number of passengers.

\section{The Boarding Process}\label{sec:boardingprocess}
We consider the boarding process from the time the passengers stand in line right outside the airplane, until the last passenger is seated. Once the passengers have entered the airplane, we assume that the queue order is maintained throughout the process and that a passenger in front cannot be passed in the aisle until that passenger has cleared the aisle and is seated. This means that most of the time the passengers stand still in the queue waiting for other passengers in front to take their seat.

The boarding is modeled as an iterative, two-step process: First, passengers move as far as they can get towards their designated row, but must stop short in the aisle if they are blocked by other passengers. This is assumed to take a negligible amount of time compared to the next step, where those passengers who stand next to their designated row use a certain aisle-clearing time to organize luggage and take a seat.

\begin{figure*}[!t]
	\begin{center}
		\includegraphics[height=6.3cm]{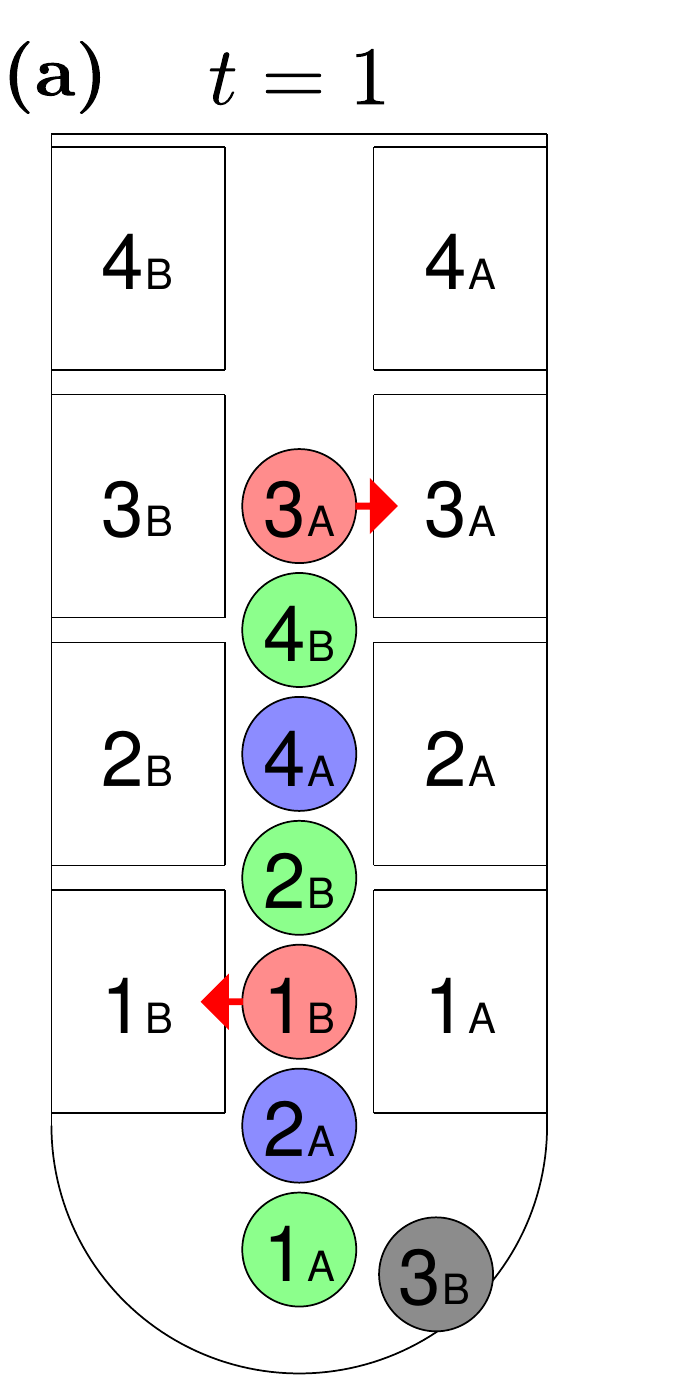}
		\includegraphics[height=6.3cm]{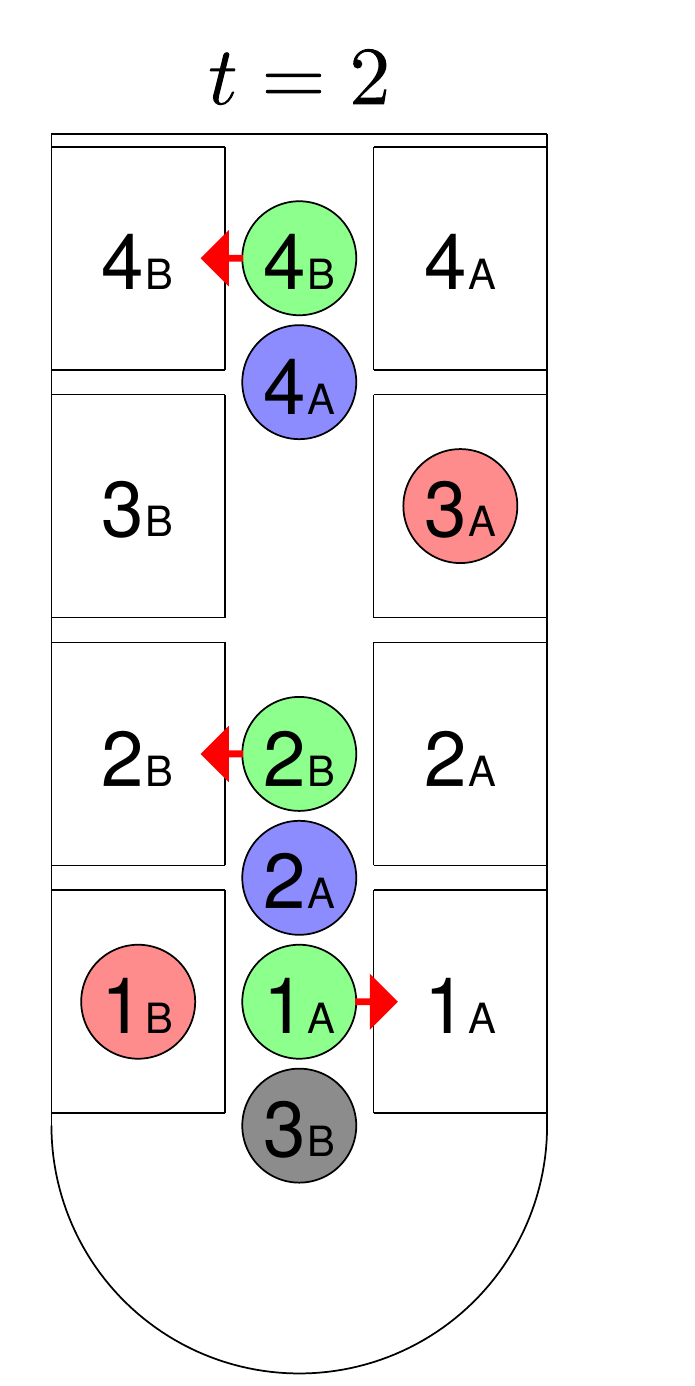}
		\includegraphics[height=6.3cm]{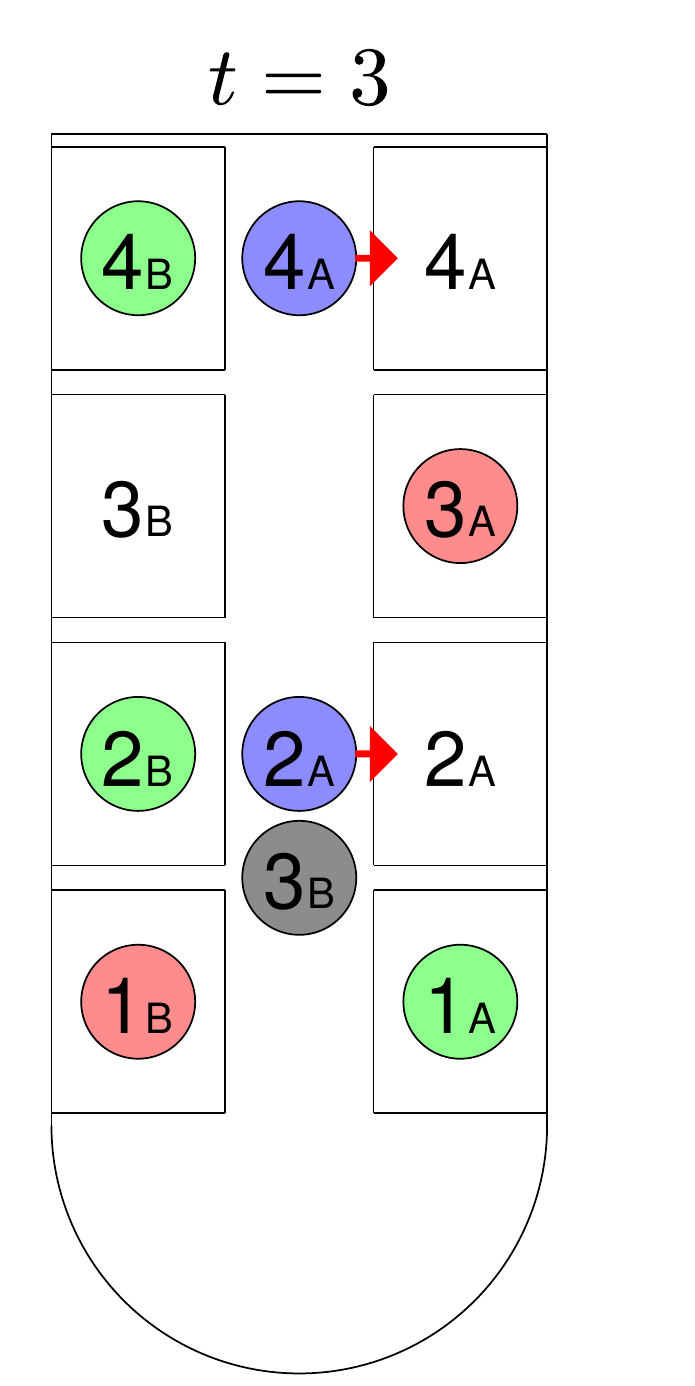}
		\includegraphics[height=6.3cm]{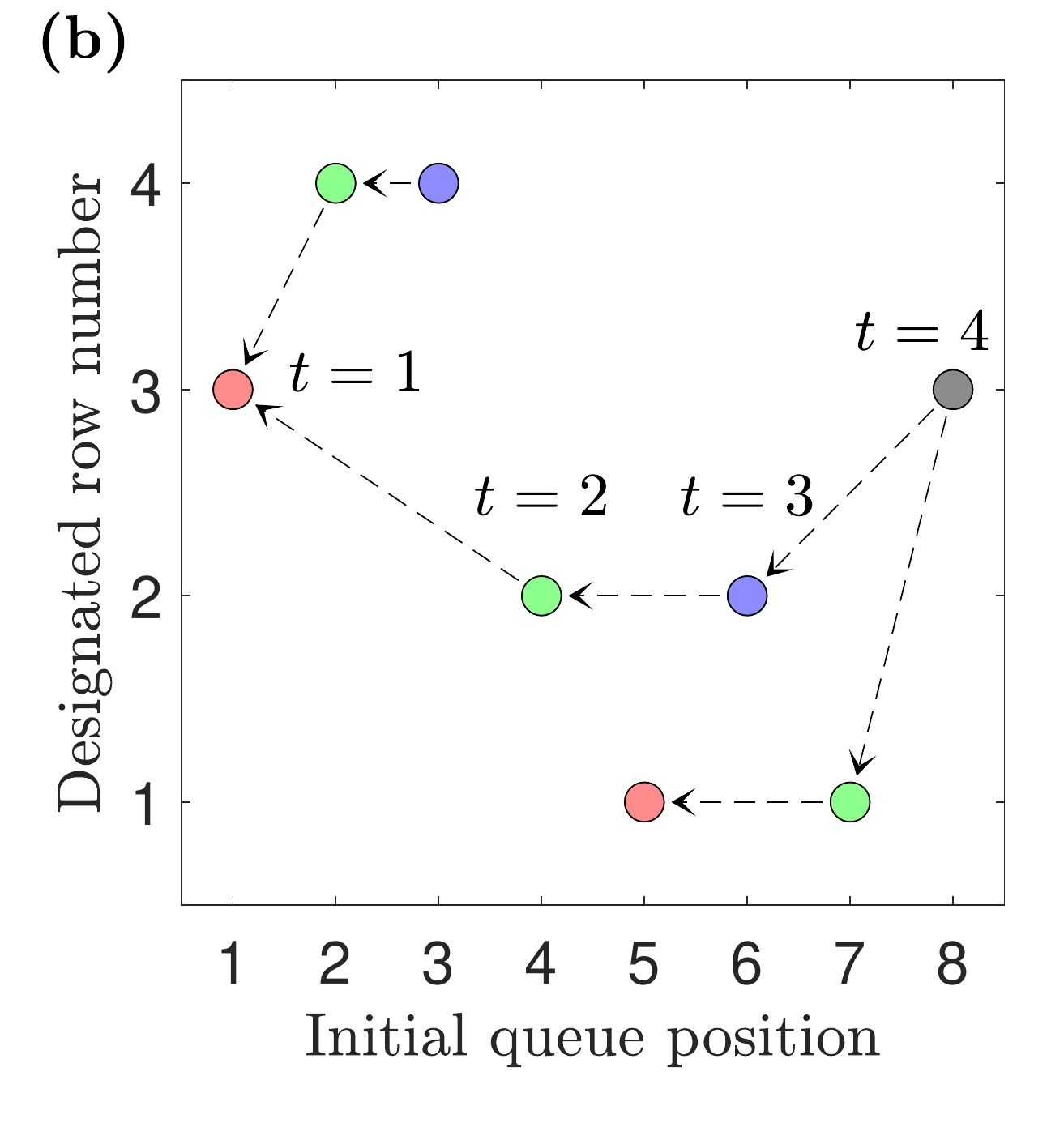}
	\end{center}	
	\caption{\label{fig:boarding_illustration} {(a)} Boarding process shown as the stepwise advance of a queue, with $N=8$ passengers, four rows and two seats per row. 
	Each circle represents a passenger with designated row number. At each time step, the queue moves forward, and passengers who have arrived at their designated rows sit down simultaneously. Passengers that take their seat at that time step are marked by red arrows, and each group of passengers that sit down simultaneously is color coded. 
	(b) Each point in the $qr$-diagram represents a passenger, indicated by initial queue position and designated row number. The color codes highlight groups of passengers that sit down simultaneously. Each passenger in a group has been blocked by at least one passenger in the preceding group, some of them indicated by arrows. Passengers that consecutively block each other form blocking chains, and the sum of aisle-clearing times of the passengers in the maximal blocking chain determines the boarding time, $T=4$.
	}
\end{figure*} 

This is illustrated by a simple example in \cref{fig:boarding_illustration}, consisting of $N=8$ passengers, all having the same aisle-clearing time. At each time step, the queue moves forward and some passengers are able to sit down. However, most passengers are delayed by other passengers blocking the aisle. 
E.g., at $t=1$ the second passenger in the queue aiming for row~4, must wait for the passenger sitting down at row~3 since the latter has a lower row number. Other passengers are blocked by displacement: the fourth passenger in the queue heading for row~2 does not have to pass any other passenger but must still wait for passengers in front to proceed. 
If there were more room for passengers to stand closer in the aisle (less congestion), the displacement effect would be less pronounced. The boarding time $T$ is the time until the last passenger is seated, and in \cref{fig:boarding_illustration}, $T=4$ time steps.

The $qr$-diagram in \crefformat{figure}{Fig.~#2#1{(b)}#3} \cref{fig:boarding_illustration} conveys each passenger's initial queue position $q$ and designated row number $r$ in the airplane. Groups of passengers that sit down simultaneously are color coded.
The boarding time can be obtained by summing up the time it takes for each group of passengers to clear the aisle.

An alternative strategy, which is used in this paper, is to utilize that each person in a group has been blocked by at least one passenger in the preceding group. The $qr$-diagram is used as a tool to analyze the entire hierarchy of blocking between passengers for a given queue and to find sequences (chains) of passengers that consecutively block each other according to a blocking relation \cite{Erland/Kaupuzs/Frette/Pugatch/Bachmat:2019}. 
The boarding time can then be found by considering the blocking chain that needs the most time to seat all its passengers. In \crefformat{figure}{Fig.~#2#1{(b)}#3} \cref{fig:boarding_illustration} three different blocking chains are shown, and the longest one consists of four passengers that have to be seated one after the other. The boarding time is the sum of those passengers' aisle-clearing times.

\section{Main Results}\label{sec:mainresults}
The boarding process is characterized by three key parameters. 
The congestion $k$ is the ratio between the length of the initial queue to the aisle length, typically in the range of $3$ to $5$ for common passenger airplanes.
The second parameter is the fraction $p$ of passengers who are considered slow, i.e., passengers with long aisle-clearing time. The remaining fraction $(1-p)$ are fast passengers.
Both the slow and the fast group have their own aisle-clearing time distribution, and each distribution is characterized by an inherent effective aisle-clearing time parameter. The ratio $C$ of the effective aisle-clearing time of the fast group to the effective aisle-clearing time of the slow group is the third parameter.

\begin{figure*}[!t]
	\begin{center}
		\includegraphics[width=10.5cm,clip]{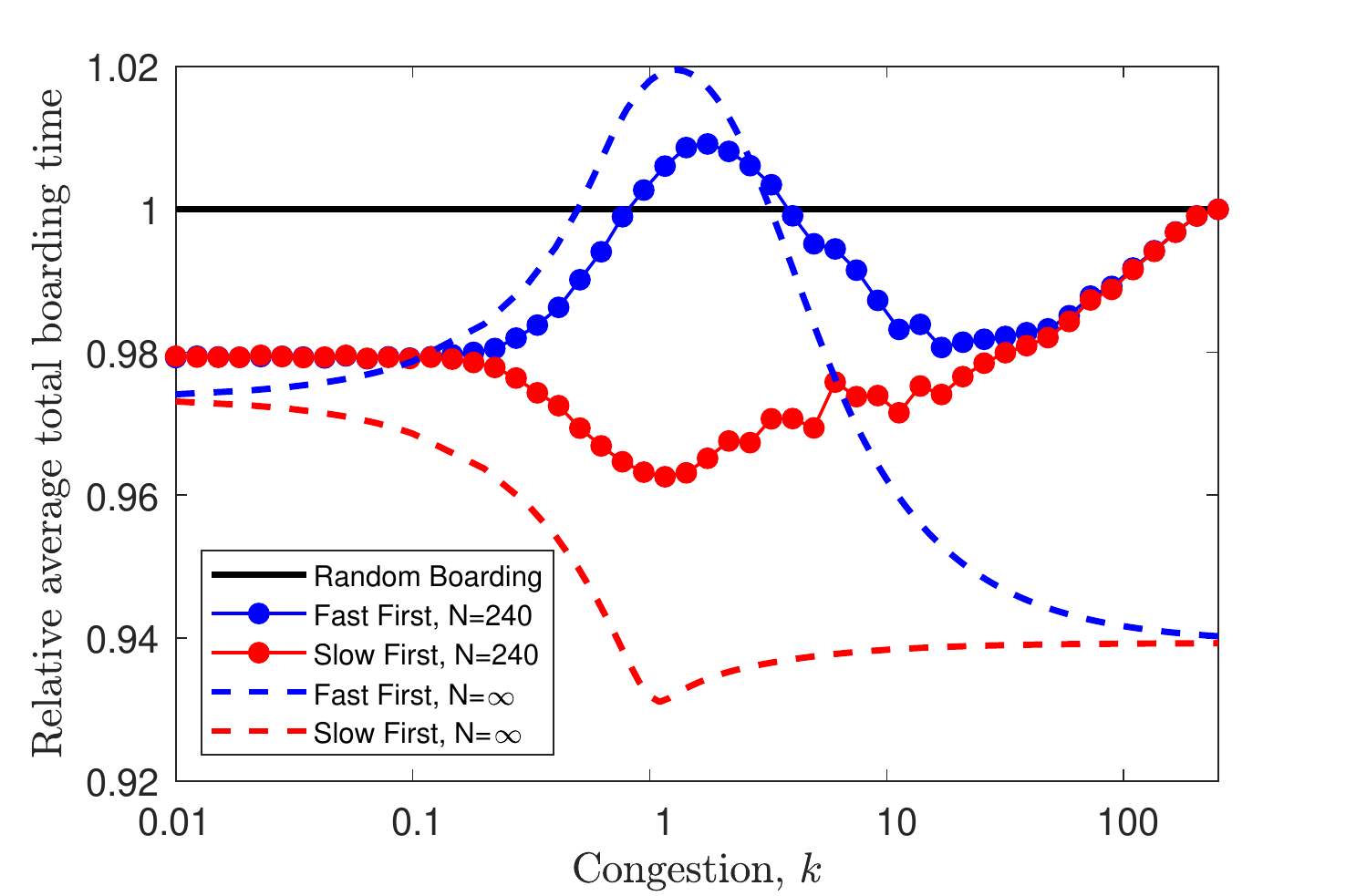}  
	\end{center}
	\caption{\label{fig:mainresults} 
	Total boarding time of the slow-first (red curves) and the fast-first (blue curves) policies relative to the random boarding policy (black curve) for increasing passenger congestion $k$. We used realistic parameters with $10\%$ slow passengers ($p=0.1$) and assumed that the fast passengers (e.g. those with little hand luggage) clear the aisle twice as fast as the slow passengers ($C=0.5$). We assumed there are 6 seats per row, and a total of $N = 240$ passengers. The results are similar for the asymptotic case with infinite number of passengers ($N=\infty$, dashed lines). 
	Remarkably, on average the slow-first policy use less time to seat all passengers than the two other policies. 
	That the slow-first policy is superior can be intuitively explained by more synchronized seating --- it better exploits the possibility to seat larger waves of passengers with similar aisle-clearing time in parallel. However, the result is non-trivial as the similar fast-first policy is inferior to the the random boarding policy for a range of congestion values $k$.
	Each point in the graph is an average of $10^6$ discrete-event runs.
	}
\end{figure*}

The main results of this paper are illustrated in Fig.~\ref{fig:mainresults}. The total boarding time of the slow-first and fast-first policies are compared to the random boarding policy for increasing congestion $k$. In the subsequent sections we prove rigorously the universal nature of the main features in Fig.~\ref{fig:mainresults} and that they apply for any set of parameters.

In Fig.~\ref{fig:mainresults}, the congestion $k$ is on the horizontal axis, the fraction of slow passengers is $p=10\%$, and the effective aisle-clearing time of the slow passengers is twice as long as for the fast passengers, i.e., $C=0.5$. 
The average total boarding time for the slow-first policy is less than for the random boarding policy for all values of $k$, both for finite number of passengers $N=240$ (red bullets) and $N\rightarrow\infty$ (red, dashed curve). However, the quite similar fast-first policy (blue curves) does not show this universal feature, as there is a range around $k\sim 1$ where fast first is inferior to random boarding. This highlights that the universal result for slow first is highly non-trivial.

When $N=240$, all three policies approach the same values as $k\rightarrow \infty$, since then each passenger would fill up the whole aisle. The seemingly erratic behavior for $k\sim 10$ is due to sudden changes when $N=240$ is a multiple of $k$.

\begin{figure*}[t!]
	\begin{center}
		\includegraphics[width=7.5cm,clip]{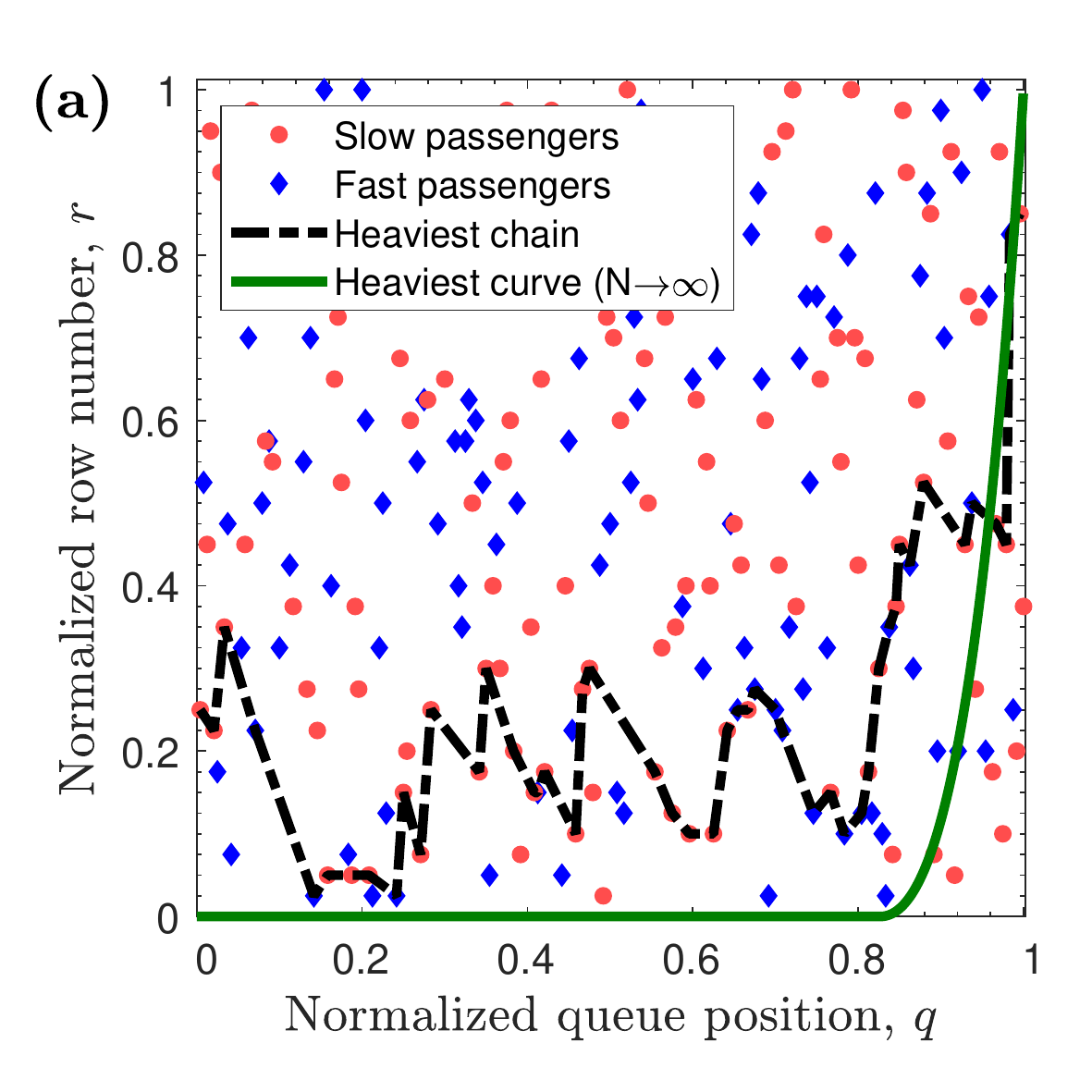}
		\includegraphics[width=7.5cm,clip]{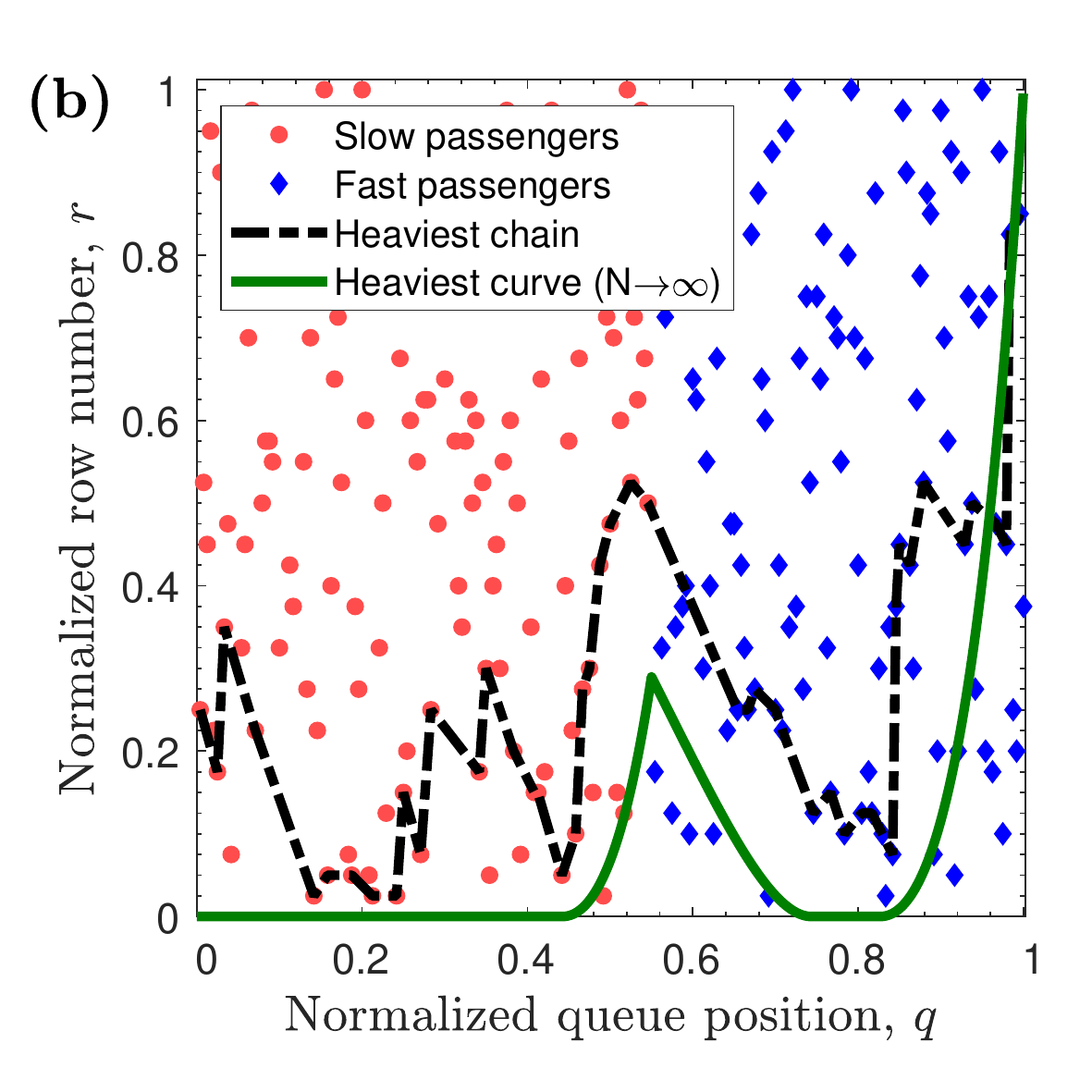}
	\end{center}
	\caption{\label{fig:policies} $qr$-diagrams for {two different} boarding policies, with each of the $N=240$ passengers marked as a point, $h=6$ seats per row, and parameters $k=4$, $p=0.55$ and $C=0.3$ derived from empirical data. 
	(a) \emph{Random boarding policy} with all passengers in one group: the passengers are uniformly distributed over the diagram; 
	(b) \emph{Slow-first policy} with two groups: the slow passengers are in the first part of the queue (red {bullets}).		
	For both policies, the boarding time is the sum of the aisle-clearing times for passengers that belong to the heaviest chain (dashed lines). 
	The preceding passenger in a chain must take his seat before the next in the chain can sit down. 
	The boarding time in each diagram is determined by the heaviest chain which follows the asymptotic limit (solid line --- the geodesic), up to statistical fluctuations that diminish for higher number of passengers.
	}
\end{figure*}

The total boarding time in Fig.~\ref{fig:mainresults} can be determined by what is called the heaviest chain of passengers. This is illustrated by the $qr$-diagrams in Fig.~\ref{fig:policies}.
The preceding passenger in a chain must take one's seat before the next in the chain can sit down. The weight of a chain is the sum of the aisle-clearing times (weights) for passengers that belong to the that chain. The boarding time is given by the weight of the heaviest chain, and that chain follows the asymptotic heaviest curve (the geodesic), up to statistical fluctuations that are diminishing for higher number of passengers ($N\rightarrow\infty$). In the asymptotic case analytical expressions exist for the total boarding time for both the random boarding and the slow-first (and fast-first) policies.
In Fig.~\ref{fig:policies}, the main parameters are derived from empirical data. More than half of the passengers are in the slow group ($p=0.55$, those with hand luggage) and their aisle-clearing time is about three times longer than for those in the fast group ($C=0.3$). The congestion is $k=4$.

The random boarding policy treats all passengers as one group with effective aisle-clearing time $\tau_A\equiv \tau_{X_A}$, 
and the boarding time depends linearly on $\tau_A$.  $\tau_A$ is determined by the distribution of the aisle-clearing times $X_A$, but analytical expressions for $\tau_A$ are generally not available. 
The distribution of $X_A$ is a mixture of the aisle-clearing time distributions of the slow and the fast groups, with respective weights $p, 1-p$ and effective aisle-clearing times $\tau_S, \tau_F$. We call the mixture \emph{concave} if \begin{equation}\label{eq:concavemixdef}
	\tau_A^2 \;\geqslant\; p\tau_S^2+(1-p)\tau_F^2 
	\;\equiv\; \hat{\tau}^2_{A,mix}.
\end{equation}
No distributions resulting in non-concave mixtures have, so far, been detected. 

Fig.~\ref{fig:mainresults} presents relative boarding times for one specific choice of the parameters $p$ and $C$. Under variations in these parameters, comparisons can conveniently be made using the relative difference 
\begin{equation*}
D(k,p,C,N)=\frac{\angles{T_\textrm{RA}}-\angles{T_\textrm{SF}}}{\angles{T_\textrm{SF}}}
\end{equation*}    
between the average boarding times of the random boarding (RA) and slow-first (SF) policies.
\begin{figure*}[t]
	\begin{center}
		\includegraphics[width=7.45cm,clip]{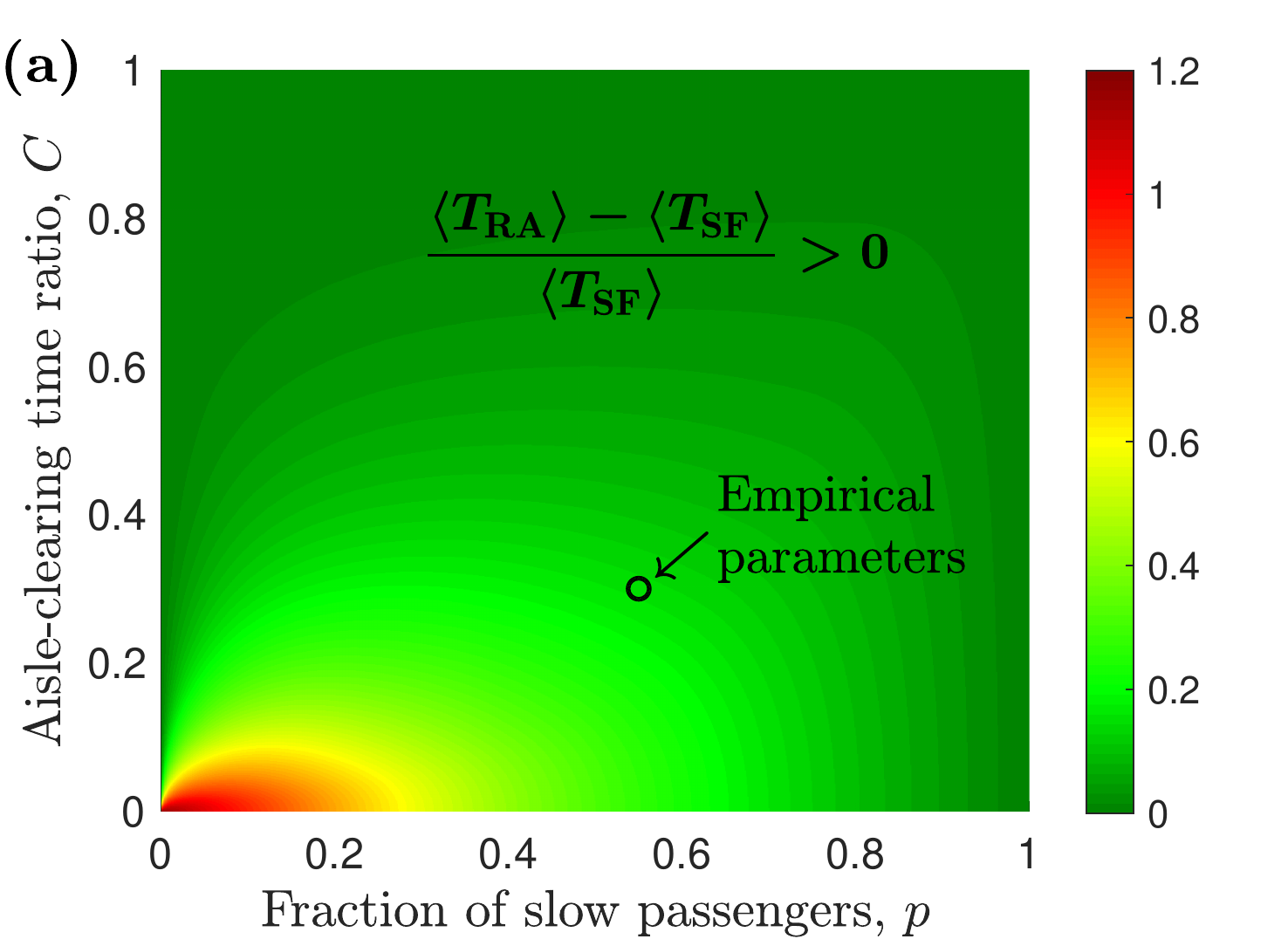}
		\includegraphics[width=8.4cm,clip]{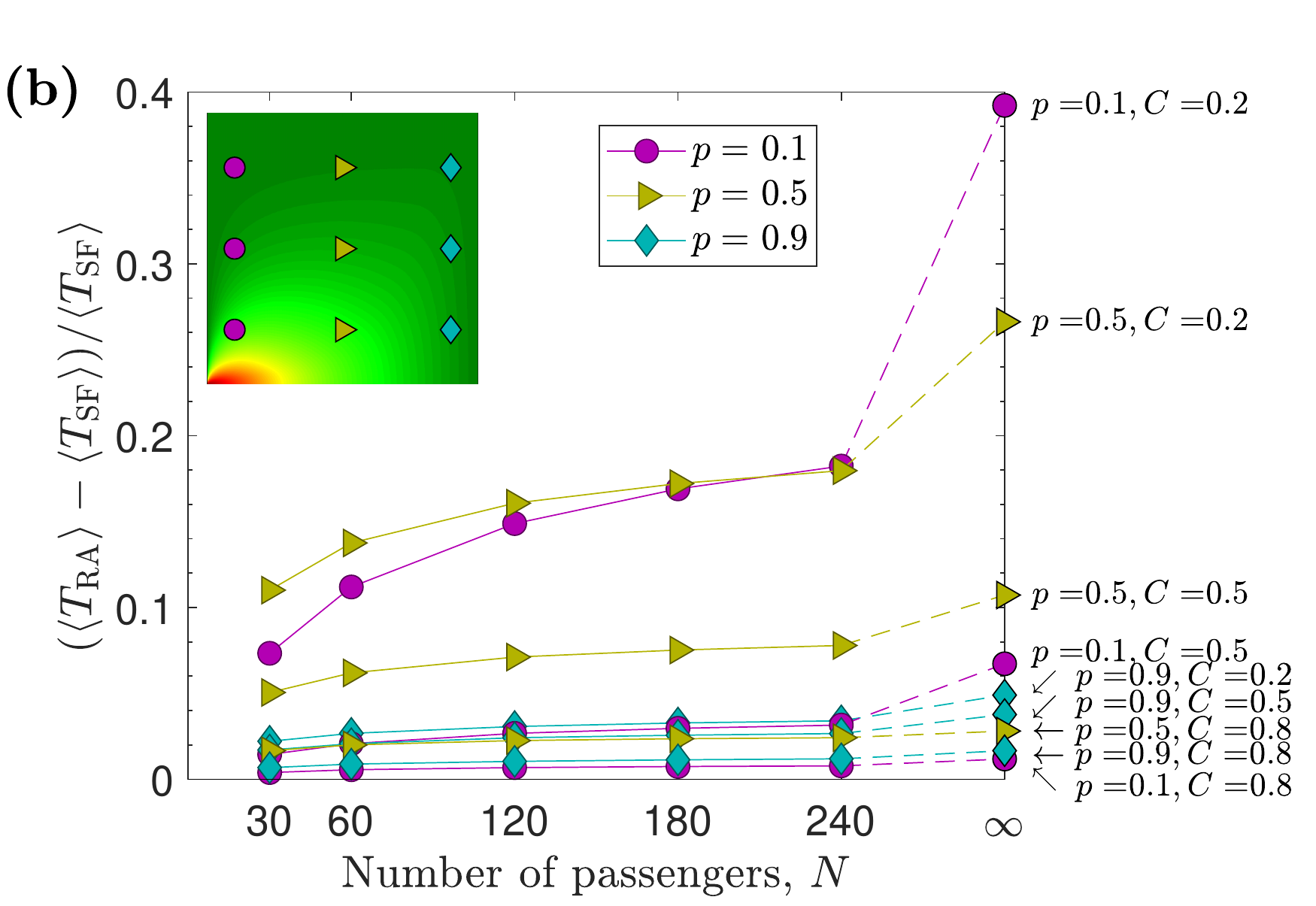}\\
	\end{center}
	\caption{\label{fig:RAvsSFcontourplot} 
	Relative difference in average total boarding time $D= (\angles{T_\textrm{RA}}-\angles{T_\textrm{SF}})/\angles{ T_\textrm{SF}}$ between the random boarding and the slow-first policies when $k=4$. 
	(a) The number of passengers $N\rightarrow \infty$, and we assume $\tau_A=\hat{\tau}_{A,mix}$. The slow-first policy gives reduced boarding time for all combinations of $p$ and $C$. The maximum relative difference of {$115\%$} is obtained for small $p$ and $C=kp/\sqrt{e^{k}-k+1}$. For the empirical parameters resulting from separating passengers into slow and fast groups based on hand luggage, the relative difference is $15\%$  (black circle). 
	(b) Simulation results for finite numbers of passenger $N$ confirm that random boarding is inferior to slow first. The connected points show that $D=(\langle T_\textrm{RA} \rangle -\langle T_\textrm{SF} \rangle)/{\langle T_\textrm{SF} \rangle}>0$ for increasing $N$ for all combinations of parameter values $p\in\{0.1,0.5,0.9\}$ and $C\in\{0.2,0.5,0.8\}$. The rightmost points are based on the asymptotic values taken from the indicated positions in the inset contour plot from (a), and corrected by precise estimates of $\tau_A$. The congestion $k=4$ with 6 seats per row, and the accuracy is $\pm 0.0002$ (as a result of $10^6$ runs for each finite-$N$ data point).
	}
\end{figure*}
When we assume that $\tau_A=\hat{\tau}_{A,mix}$ in \cref{eq:concavemixdef}, the contour plot in \crefformat{figure}{Fig.~#2#1{(a)}#3}\cref{fig:RAvsSFcontourplot} shows that $D>0$ in the $(p,C)$ unit square for $k=4$ and $N\rightarrow\infty$. Since $\angles{T_\textrm{RA}}$ is scaled by $\tau_A$, it follows that $D>0$ also holds for $\tau_A>\hat{\tau}_{A,mix}$ (all concave mixtures).

Our main result can be stated as follows.
\begin{theorem}\label{th:RAvsSF}
	The expected boarding time $\angles{T}$ is shorter for the slow-first policy than for the random boarding policy for all values of $k>0$ and $p,C \in (0,1)$, in the asymptotic regime when $N\rightarrow\infty$, if and only if the aisle-clearing time mixture is concave. 
\end{theorem}
The discrete-event simulation results in \crefformat{figure}{Fig.~#2#1{(b)}#3}\cref{fig:RAvsSFcontourplot} indicate that this universal result also holds for realistic numbers of passengers $N$. For the empirical parameters resulting from separating passengers into slow and fast groups based on hand luggage, 
the average boarding time is $13\%$ longer with random boarding compared to slow first when $N=240$. 
	
We also show that the relative difference $D$ between random boarding and slow first can be maximized for fixed $k$ in the asymptotic regime when $\tau_A=\hat{\tau}_{A,mix}$ and $N\rightarrow\infty$. For $k>\ln(2)$ the maximum is obtained for $p\rightarrow 0$ and $C=kp/\sqrt{e^k-k+1}$
\begin{align}\label{eq:RAvsFFmaxmain}
\sup_{p,C} D & \geq \frac{k-\ln(2)+1}{\sqrt{k}} - 1  
\quad \overset{k\rightarrow\infty}{\sim} \quad \sqrt{k}-1.
\end{align}

For $k=4$, $\sup D= 115\%$ when $\tau_A=\hat{\tau}_{A,mix}$. For concave mixtures where $\tau_A>\hat{\tau}_{A,mix}$, the maximum relative difference is even higher. For large $k$ it is obviously unbounded, however, for fixed $p,C$ the asymptotic relative difference approach a limiting value when $k \rightarrow\infty$. 
Proofs for \cref{th:RAvsSF} and \cref{eq:RAvsFFmaxmain} appear in Appendix \ref{app:proofs}. 

\clearpage
\section{Boarding process and asymptotic boarding time}\label{sec:model}
In this section we give a brief introduction to how the boarding time can be approximated by the weight of the heaviest chain as $N\rightarrow\infty$. The reader is referred to Ref. \cite{Erland/Kaupuzs/Frette/Pugatch/Bachmat:2019} for a simple introduction to the analogy between airplane boarding and spacetime geometry and to Ref. \cite{Bachmat:2014} for a more rigorous mathematical description. 

\subsection{Main parameters}\label{ssec:mainparameters}
{The following parameters govern the boarding process.} 

(i) \emph{The total number of passengers $N$}. For simplicity, we assume a full airplane, i.e., that $N$ equals the number of seats in the airplane. In Fig.~\ref{fig:boarding_illustration}, $N=8$.

(ii) \emph{Congestion $k$, the length of the queue before boarding relative to the length of the aisle.} 
Let $h$ be the number of seats per row, $d$ the distance between consecutive rows and $w$ the distance between passengers {as they are lined up in the aisle}. Then $k=hw/d$. The parameter $k$ reflects both the maximum density of passengers {in the queue} and the interior design of the airplane.\footnote{The parameter $k$ can also be generalized to include the number of aisles and the relative occupancy of the airplane \cite{Bachmat/Berend/Sapir/Skiena/Stolyarov:2009}.}  The value $k=1$ is used for simplicity in Fig.~\ref{fig:boarding_illustration}.

(iii) \emph{Fraction of slow passengers $p$.} In Fig.~\ref{fig:boarding_illustration}, the aisle-clearing time of all passengers is the same, so $p=1$.

(iv) \emph{Effective aisle-clearing time $\tau_X$.} The aisle-clearing time $X$ is the time needed for a passenger to organize bin luggage and take a seat after reaching one's row. In Fig.~\ref{fig:boarding_illustration} all passengers have an aisle-clearing time of $X\equiv 1$ time steps. The effective aisle-clearing time $\tau_X$ is an intrinsic parameter in the asymptotic estimate of the total boarding time and is determined by the distribution of $X$ only. Explicit expressions are generally not available, except when $X\equiv c$ is constant, in which case $\tau_X=c$.  

(v) \emph{Ratio $C$ between the effective aisle-clearing time of the fast passengers to that of the slow passengers.}  

(vi) \emph{Initial queue position $q$ of a passenger normalized by the total number of passengers $N$}. In Fig.~\ref{fig:boarding_illustration}, the fourth passenger in the queue (heading for row 2) has $q=4/8=0.5$.

(vii) \emph{Designated row number $r$ for a passenger normalized by the total number of rows}. The fourth passenger in the queue in Fig.~\ref{fig:boarding_illustration} has $r=2/4=0.5$. 

In Fig.~\ref{fig:boarding_illustration} the actual queue and row numbers are given on the axes, while the normalized $(q,r)$-values will be used in the remaining part of the paper.

\subsection{Boarding policies and heaviest blocking chains}\label{ssec:boardingpolicies}
A boarding policy is the way the queue of passengers is organized. The most common policy is the unorganized random boarding policy, where passengers enter the queue in random order. A typical scenario with the random boarding policy is illustrated in the $qr$-diagram in \crefformat{figure}{Fig.~#2#1{(a)}#3}\cref{fig:policies}. The points representing each of the $N=240$ passengers are uniformly distributed over the unit square.
A scenario with the slow-first policy is shown in \crefformat{figure}{Fig.~#2#1{(b)}#3}\cref{fig:policies}. The passengers are divided into two groups based on aisle-clearing time, and those who are considered slow constitute the first part of the queue.  Within each group, the passengers are randomly distributed in the queue and so are the designated row numbers.

The task of finding the total boarding time can be found by identifying the groups of passengers (wave-fronts) that sit down simultaneously as shown in Fig.~\ref{fig:boarding_illustration} where all passengers have equal aisle-clearing time. In this case the boarding time can be found by counting the number of such wave-fronts and multiply by the aisle-clearing time. 

An important additional feature that can be observed in Fig.~\ref{fig:boarding_illustration} is that each passenger in a wave-front has been blocked from sitting down any earlier by a passenger in the preceding wave-front. This feature is particularly useful when the aisle-clearing times vary between passengers. Instead of laboriously counting wave-fronts, the blocking hierarchy between passengers is exploited. We say that a passenger $A$ \emph{blocks} passenger $B$ if $A$ must clear the aisle before $B$ can take a seat. A blocking chain consists of passengers that consecutively block each other.  
The weight of a blocking chain is the sum of aisle-clearing times for passengers that belong to that chain, and the crucial observation is that the boarding time equals the weight of the heaviest chain. 

Given a queue, we can construct a heaviest chain by starting with a passenger $B$ in the last wave-front. Several passengers in the preceding wave-fronts may be blocking $B$. The one that is closest in the queue just before $B$ arrives at $B$'s designated row is chosen as the next passenger in the chain. A heaviest chain is obtained by proceeding in this manner, until reaching a passenger near the front of the queue, who was never blocked. Examples of heaviest chains are shown in \crefformat{figure}{Fig.~#2#1{(b)}#3} \cref{fig:boarding_illustration} and Fig.~\ref{fig:policies} for $N=8$ and $N=240$, respectively. {A much more efficient way of constructing the heaviest blocking chain is given in \cite{Jacobson/Vo:1992} for $k=0$, and we apply this to compute $\tau_X$ in \cref{sec:CX}.}

\subsection{Space-time geometry, causal curves and curve weight}\label{ssec:blockingchains}
Fig.~\ref{fig:policies} indicates that the heaviest chains are close to what is called a heaviest curve. When the number of passengers  $N\rightarrow\infty$, tools from causal set theory and space-time geometry can be used to show that the heaviest chains do approach the heaviest curve in the limit \cite{Bachmat:2014}. 

In space-time geometry a Lorentzian metric defines whether there is a causal relation between events (time-like separation), and the proper time between such events are measured by causal curves with maximal length under that metric (geodesics). 
In airplane boarding passengers correspond to events and the property that passenger $A$ blocks passenger $B$ corresponds to that of $A$ potentially having a casual influence on $B$. 
The time between seating of two passengers corresponds to the proper time between events, and when the number of events (passengers) $N\rightarrow\infty$, the time difference can be calculated by a particular Lorentzian metric.

The total boarding time is given by the blocking chain with a maximal sum of aisle-clearing times, and this heaviest blocking chain approaches a heaviest causal curve under the metric when $N\rightarrow \infty$. If we assume that the aisle-clearing time of each passenger is deterministically given by the passengers' normalized queue and row position, $X=\tau(q,r)$, the \emph{curve weight} of a causal curve $r(q)$ can then be defined by \cite{Bachmat:2014}
\begin{equation}\label{eq:weight}
W(r) = \int_{q_0}^{q_1} \tau(q,r(q)) \sqrt{r'(q) + k[1-r(q)]}dq.
\end{equation}
The square root in the integral measures the number of passengers along the causal curve, and this is weighted by the respective aisle-clearing time $\tau(q,r(q))$ along the same part of the curve. The curve is causal when the square root is real, and since the normalized queue and row number $(q,r)$ of all passengers are within the unit square, the heaviest curve must obey the same restriction. Other features of the heaviest curve is that it connects the points $(0,0)$ and $(1,1)$, it is continuous, and it is differentiable whenever $\tau(q,r)$ is continuous \cite{Bachmat:2014}. 

As noted in \cref{sec:mainresults}, when the aisle-clearing times $X$ are stochastic, 
$\tau(q,r)$ in \cref{eq:weight} can be replaced by an effective aisle-clearing time $\tau_X(q,r)$, which is independent of $k$ \cite{Bachmat:2019}. 
Moreover, we only consider boarding policies where each section of the queue (each passenger group) has the same aisle-clearing time distribution, such that $\tau_X(q,r)=\tau_X(q)$ take constant values on a finite number of $q$-intervals.

\subsection{Asymptotic boarding time}\label{sec:analysis}
A general formula of Myrheim \cite{myrheim:1978} states that the total boarding time converges to a multiple of the weight (\cref{eq:weight}) of the heaviest causal curve $r$ from $(0,0)$ to $(1,1)$ within the unit square \cite{Bachmat:2014},
\begin{equation*}
\frac{T}{\sqrt{N}} \;\overset{\textrm{a.s.}}{\to}\; 2 \max_{r}W(r).
\end{equation*}
From this, the asymptotic average boarding time is given by
\begin{equation}\label{eq:Tapprox}
\angles{T} \;\sim\; 2\sqrt{N} \max_{r}W(r)  \;\equiv\; \hat{T}.
\end{equation}
The asymptotic boarding time $\hat{T}$ is a leading term, and tends to over-estimate the finite-$N$ average boarding time $\langle T\rangle$ by a relative error of order $O(N^{-\frac{1}{3}})$ for $k\leqslant \ln(2)$ and  $o(N^{-\frac{1}{4}})$ for $k> \ln(2)$ \cite{Bachmat/Khachaturov/Kuperman:2013}. Still, the relative ranking of boarding policies is usually maintained for small $N$. 

For the simple case with $k=0$ and aisle-clearing time $X\equiv1$, \cref{eq:Tapprox} reduces to the Vershik-Kerov theorem which states that the number of passengers in a maximal blocking chain is $2\sqrt{N}$ for large $N$ \cite{Vershik/Kerov:1977}. Extending to the case where all passengers have stochastic aisle-clearing times $X$ from the same distribution, $\angles{T} \;\sim\; 2\tau_X \sqrt{N}$ \cite{Bachmat:2014}. These relations are used to estimate and establish bounds for $\tau_X$ in \cref{sec:CX}. 

The result in \cref{eq:Tapprox} is used in \cref{sec:RAvsSF} to derive analytical expressions for the expected boarding time for both the random boarding and the slow-first boarding policies.

\section{Effective aisle-clearing time and empirical data}\label{sec:CX}
The effective aisle-clearing time $\tau_X(q)$, replaces $\tau(q,r)$ in \cref{eq:weight} when each section of the queue consists of groups of passengers with the same aisle clearing time distribution. $\tau_X$ for each group depends solely on the distribution of $X$ and is independent of $k$. Explicit analytical expressions for $\tau_X$ are to our knowledge only known 
for deterministic $X\equiv c$, in which case $\tau_X=c$. However, the analytical bounds and extensive numerical calculations in this section indicate that $\sqrt{\langle X^2 \rangle}$ is a relatively tight lower bound for $\tau_X$. The stronger conjecture, that a mixture of two distributions is concave (\cref{eq:concavemixdef}), is supported by simulations based on empirical aisle-clearing time data.

\subsection{Lower and upper bounds on the effective aisle-clearing time}
Bounds for the effective aisle-clearing time $\tau_X$ 
appear in \cite{Bachmat:2019}. Let the aisle-clearing times $X$ follow a distribution with density function $f(x)$ and support in the interval $[a,b]$. Then the following inequalities hold,
\begin{equation}\label{eq:tauejorbounds}
\frac{1}{e\sqrt{1+\ln(b/a)}} 
\leqslant \frac{\tau_X}{\sqrt{\langle X^2 \rangle}}
\leqslant {e\sqrt{1+\ln(b/a)}}. 
\end{equation}
These bounds justify the use of  $\sqrt{\langle X^2 \rangle}$ as an estimate for $\tau_X$. Moreover, they show that the ratio of upper and lower bounds of $\tau_X$ for any distribution with support in $[a,b]$ is at most $e^2[\ln (b/a)+1]$, and therefore reasonably good. Here we improve those bounds and show that the estimate $\sqrt{\langle X^2 \rangle}$ has the same bounds. 
\begin{theorem}\label{th:taubounds}
	Let the aisle-clearing times $X$ follow a distribution with density function $f(x)$ and support in the interval $[a,b]$. Then the following inequalities hold,
	\begin{equation}\label{eq:taubounds}
	\max_{u\in [a,b]} \frac{\int_u^btf(t)dt}{\sqrt{\int_u^bf(t)dt}}
	\;\leqslant\; {\tau_X}
	\;\leqslant\; \min_{\{u_i\}} \sum_{i=1}^n u_i\sqrt{p_i} \;\equiv\; \tau_{X}^{U}, 
	\end{equation}
	where $\{u_i\}$ is a sub-division of $[a,b]$ with $a=u_0< u_1<u_2,\ldots <u_n=b$ and $p_i=Pr(u_{i-1}< X \leqslant u_i)$. The upper and lower bounds for $\tau_X$ also holds for $\sqrt{\langle X^2 \rangle}$.
\end{theorem}
The proof is presented in \cref{apps:taubounds}.
Based on the extensive numerical calculations in Sec. \ref{ssec:jevgenijs} below, we conjecture that $\tau_X^2 \geqslant \langle X^2 \rangle$ for all distributions of $X$. This is necessary for all mixtures to be concave, which in turn is a prerequisite for \cref{th:RAvsSF} to hold.

For a two-valued distribution taking the values $X=\tau_S$ and $X=\tau_F$ with probabilities $p$ and $1-p$, respectively, the ratio of the upper bound in \cref{eq:taubounds} and the conjectured lower bound is
\begin{equation}\label{eq:tauratio}
  	\frac{\tau_{X}^{U}}{\sqrt{\langle X^2 \rangle}}
  	= \sqrt{1 + \frac{2C\sqrt{p(1-p)}}{p+C^2(1-p)}}.
\end{equation}
This ratio attains the maximum value $\sqrt{2}$ when $p=C^2/(1+C^2)$ for all $C\in (0,1)$.

\subsection{Numerical estimation of $\tau_X$ for two-valued distributions}\label{ssec:jevgenijs}
In this section we base the numerical estimation of the effective aisle-clearing time $\tau_X$ on the asymptotic behavior of the normalized boarding time when $k=0$,
\begin{equation}\label{eq:tau_asymptotics}
	\frac{\angles{T}}{\sqrt{N}} \sim 2\tau_X - a_1 N^{-\frac{1}{3}},
\end{equation}
where the constant $a_1>0$ \cite{Bachmat/Berend/Sapir/Skiena/Stolyarov:2006}.

The slow convergence towards $\tau_X$ in \cref{eq:tau_asymptotics} requires large $N$ to obtain precise estimates of $\tau_X$. However, the computation of $\angles{T}$ is challenging when the number of passengers $N$ is large, in particular if a discrete event simulation approach is used to compute $T$ for every realization of a queue of passengers. 
	
We describe a more efficient procedure for the estimation of $\tau_X$. In particular, we apply an efficient algorithm for the computation of the boarding time $T$ for a given queue. In a more general setting this is the same as computing the heaviest increasing subsequence in a permutation (HIS) \cite{Jacobson/Vo:1992} which can be done in time of order $N \ln N$. The procedure is described in more detail in \cref{app:HIS}. E.g., we estimated the boarding time $T$ in 9 minutes running time for $N$ as large as $2.62 \cdot 10^8$. 
For comparison, a discrete event simulation would be of order $N^{3/2}$ in the {simplest} case, where all passengers are equally fast. Indeed, in this case the boarding proceeds in $\sim \sqrt{N}$ time steps, and it is necessary to update positions of $\sim N$ passengers in one step. 

For two-valued distributions of the aisle-clearing time $X$ and fixed $p$, the estimates of $\tau^2_X/{\langle X^2 \rangle}$ are presented in \cref{tab:tau_twovalued}. The fact that no values are (significantly) below 1, seems to indicate that $\sqrt{\langle X^2 \rangle}$ might be a lower bound for $\tau_X$. Hence, the two-valued mixtures in the table can with reasonable certainty be said to be concave.

\begin{table}[h]
	\caption{\label{tab:tau_twovalued} 
		$\tau_X/\sqrt{\langle X^2 \rangle}$ for various combinations of $C\equiv\tau_F/\tau_S$ (numbers in the top row) and $p$ (numbers in the left column) for a two-valued distribution. 
		The precise estimates of the effective aisle-clearing time parameters $\tau_X$ have been obtained by extensive simulations.   
	}	
	\begin{ruledtabular}
		\begin{tabular}{lccccccc}
			$p$ $\backslash$ $C$& 0.01 & 0.03 & 0.1 & 0.3 & 0.5 & 0.7 & 0.9 \\
			\colrule
			0.00001 & 1.1240(20) & 1.02662(52) & 1.00127(17) & & & & \\
			0.00003 & 1.2049(21) & 1.05731(58) & 1.00401(20) & 1.00000(16) & & & \\
			0.0001  & 1.2643(23) & 1.11353(79) & 1.01211(19) & 1.00038(14) & 0.99999(14) & & \\
			0.0003  & 1.2385(19) & 1.18219(91) & 1.02929(20) & 1.00119(11) & 1.00015(11) & {1.00014(11)} & 0.99993(11) \\
			0.001   & 1.1646(14) & 1.2252(11)  & 1.06751(38) & 1.00365(17) & 1.00025(16) & 0.99995(16) & 0.99994(16) \\
			0.003   & 1.1028(11) & 1.20197(80) & 1.12042(45) & 1.01041(18) & 1.00155(16) & 1.00073(16) & {1.00042(16)} \\
			0.01   & 1.05402(69) & 1.13153(59) & 1.15914(47) & 1.02803(20) & 1.00462(17) & 1.00171(16) & 1.00054(16) \\
			0.03   & 1.02896(51) & 1.07480(43) & 1.14159(42) & 1.05433(23) & 1.01187(17) & 1.00496(16) & 1.00153(16) \\
			0.1    & 1.01313(33) & 1.03446(29) & 1.08511(32) & 1.07420(23) & 1.02711(18) & 1.01215(16) & 1.00389(16) \\
			0.3    & 1.00455(23) & 1.01376(21) & 1.03753(23) & 1.05582(21) & 1.03520(18) & 1.02031(17) & 1.00646(16) \\
			0.5    & 1.00232(20) & 1.00715(17) & 1.02042(20) & 1.03617(19) & 1.02858(17) & 1.01999(16) & 1.00735(16) \\
			0.7    & 1.00142(17) & 1.00374(15) & 1.01057(18) & 1.02050(17) & 1.01899(17) & 1.01443(16) & 1.00582(16) \\
			0.9    & 1.00051(17) & 1.00092(14) & 1.00334(16) & 1.00710(16) & 1.00779(16) & 1.00637(16) & 1.00265(16) \\
			0.97  & 1.000043(54) & 1.00022(14) & 1.00081(16) & 1.00209(16) & 1.00257(16) & 1.00194(16) & 1.00100(16) \\
		\end{tabular}
	\end{ruledtabular}
\end{table}

For each fixed $C$, the maximum of $\tau^2_X/{\langle X^2 \rangle}$ in \cref{tab:tau_twovalued} is obtained when $p\approx C^2/(1+C^2)$, the same relation that gives the maximum for the upper bound in \cref{eq:tauratio}. For the smallest value $C=0.01$, $\tau_X$ is 26\% larger than $\sqrt{\langle X^2 \rangle}$, which is relatively close to the upper limit of 41\% in \cref{eq:tauratio}. For $C\rightarrow 1$, the distribution of $X$ approaches a constant, which implies that $\tau_X \rightarrow \sqrt{\langle X^2 \rangle}$. Hence, the ratio in the table approaches 1, which makes the upper bound artificially high for values of $C$ in this range.

\subsection{$\tau_X$ for slow and fast groups based on empirical aisle-clearing time data}\label{ssec:empirical}
Passengers can be separated into a slow and a fast group in several ways depending on what information is known for each passenger. 
{Based on the empirical data in \cite{Steiner:2008}, we apply a strategy for separating slow and fast passengers into two groups based on the number of items each passenger carries:} E.g., a passenger is in the slow group if he carries more than $i$ items, where $i$ is a chosen value. In the empirical data in \cite{Steiner:2008}, the number of hand luggage items carried by each passenger proved to be the parameter which influenced aisle-clearing time the most.

The parameters resulting from this type of luggage-based separation strategy are presented in \cref{tab:tau_Steiner}, and details of computation are found in \cref{app:HIS}. Notice that the given mixtures are concave for each type of separation. This means that the main result in \cref{th:RAvsSF} applies (for $N\rightarrow\infty$). \cref{fig:RAvsSFcontourplot} confirms that slow first outperforms random boarding also for a finite and realistic number of passengers $N$ when the slow group consists of those who carry one or more hand luggage.

\begin{table}[h]
	\caption{\label{tab:tau_Steiner} Effective aisle-clearing times for groups resulting from different strategies for separating passengers into slow and fast group. The random boarding policy with all passengers in one group has exact value $\tau_A= 0.443$, while $\sqrt{\langle X_A^2 \rangle}=0.397$ (both rounded to three significant digits). All mixtures are concave since $\tau_A^2 \;\geqslant\; p\tau_S^2+(1-p)\tau_F^2 \equiv \hat{\tau}^2_{A,mix}$.}
	\begin{ruledtabular}
		\begin{tabular}{lccccccc}
			\textrm{Slow group}&
			$p$&
			$\sqrt{\langle X_S^2 \rangle}$&
			$\tau_S$&
			$\sqrt{\langle X_F^2 \rangle}$&
			$\tau_F$&
			$\hat{\tau}_{A,mix}$&
			$C\equiv \tau_F/\tau_S$\\
			\colrule
			All passengers      & 1    & 0.397 & 0.44320(12) &   -   &  	-  &  	-  &  -   \\
			$\geqslant 1$ item & 0.55 & 0.516 & 0.56756(13) & 0.153 & 0.169853(43) & 0.437 & 0.299\\
			$\geqslant 2$ items & 0.15 & 0.766 & 0.82599(16) & 0.285 & 0.317433(83) & 0.435 & 0.384\\
			$\geqslant 3$ items & 0.04 & 1.054 & 1.12911(21) & 0.348 & 0.383802(94) & 0.435 & 0.340\\
			\colrule
			$0$ or $\geqslant 2$ items 		& 0.60 & 0.408 & 0.45560(13) & 0.381 & 0.41827(10) & 0.441 & 0.918\\
			$2$ items 			& 0.11 & 0.645 & 0.68426(12) & 0.352 & 0.39959(12) & 0.442 & 0.584\\
			\colrule
			$1$ or $2$ items 		& 0.52 & 0.453 & 0.49250(10) & 0.327 & 0.37614(13) & 0.440 & 0.764\\
			$1$ or $\geqslant 3$ items 		& 0.44 & 0.476 & 0.53154(14) & 0.322 & 0.356265(92) & 0.442 & 0.670\\
		\end{tabular}
	\end{ruledtabular}
\end{table}


\clearpage
\section{Analysis of the random boarding and slow-first policies}\label{sec:RAvsSF}
We now turn to  computing the asymptotic average boarding time in \cref{eq:Tapprox} for $k>0$, {with} $p$, $C\equiv \tau_F/\tau_S$ both in the range $(0,1)$. We show that the slow-first policy is more efficient than the random boarding policy in the entire $(k,p,C)$-parameter space in the large-$N$ limit ($N\rightarrow \infty$) when the aisle clearing time mixture is concave. Comparisons to simulation results for finite $N$ are also made.

\subsection{Analysis of the random boarding policy}\label{ssec:Tcompute_RA}
The random boarding policy treats all passengers as one group such that all passengers have the same aisle-clearing time distribution and effective aisle-clearing time. Then $\tau(q,r)$ can be replaced by $\tau_A$ in \cref{eq:weight} such that $W(r)=\tau_A L(r)$, where
\begin{equation}\label{eq:length}
	L(r) = \int_{q_0}^{q_1} \sqrt{r'(q) + k[1-r(q)]} \, dq.
\end{equation}
The variational method can be used to maximize the curve length $L(r)$, and this leads to general solutions of the form $r^*(q)=ae^{2kq}+be^{kq}+1$ when $k>0$.

\begin{figure*}[htb]
	\begin{center}
		\includegraphics[width=4.0cm,clip]{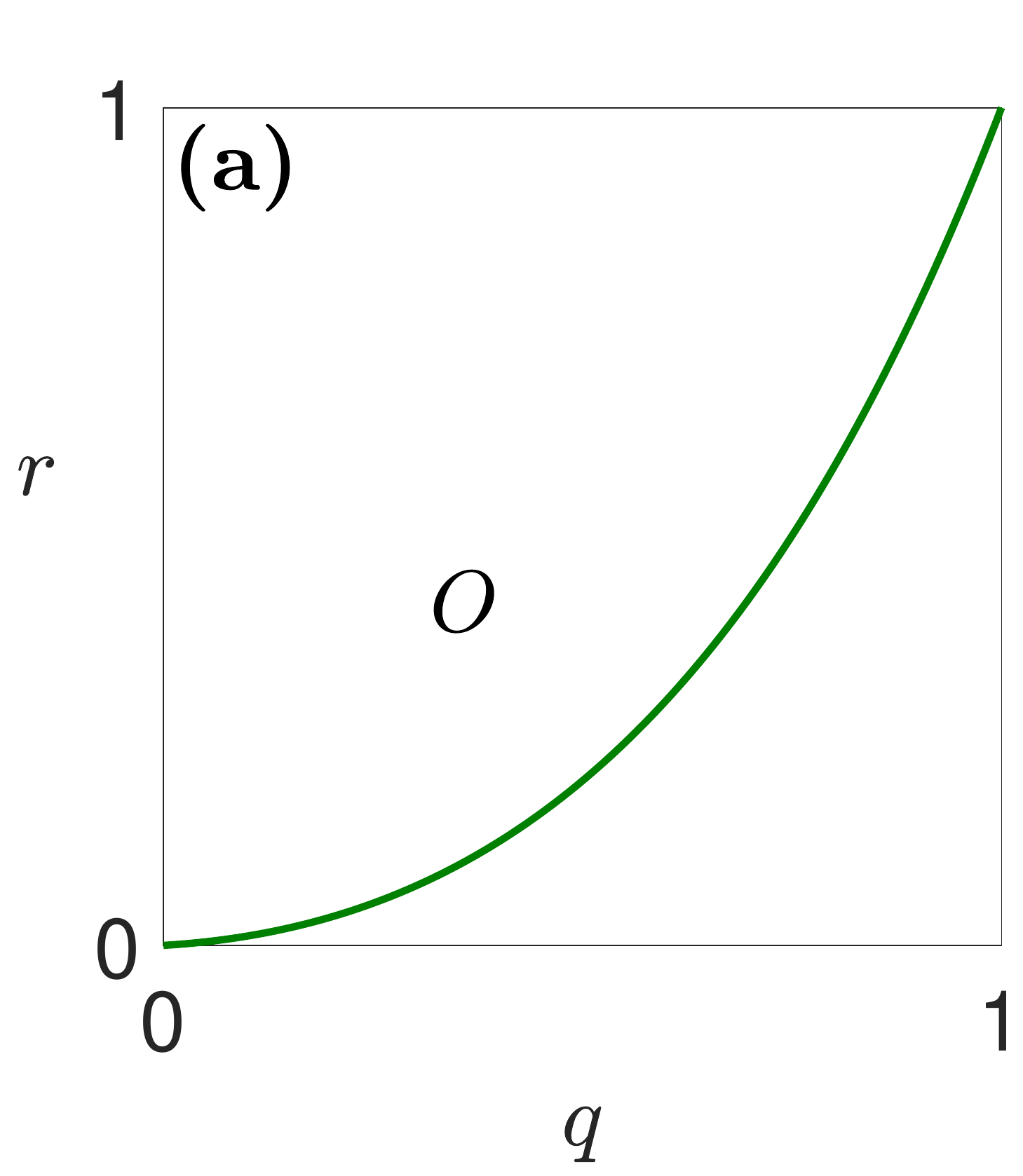}
		\includegraphics[width=4.0cm,clip]{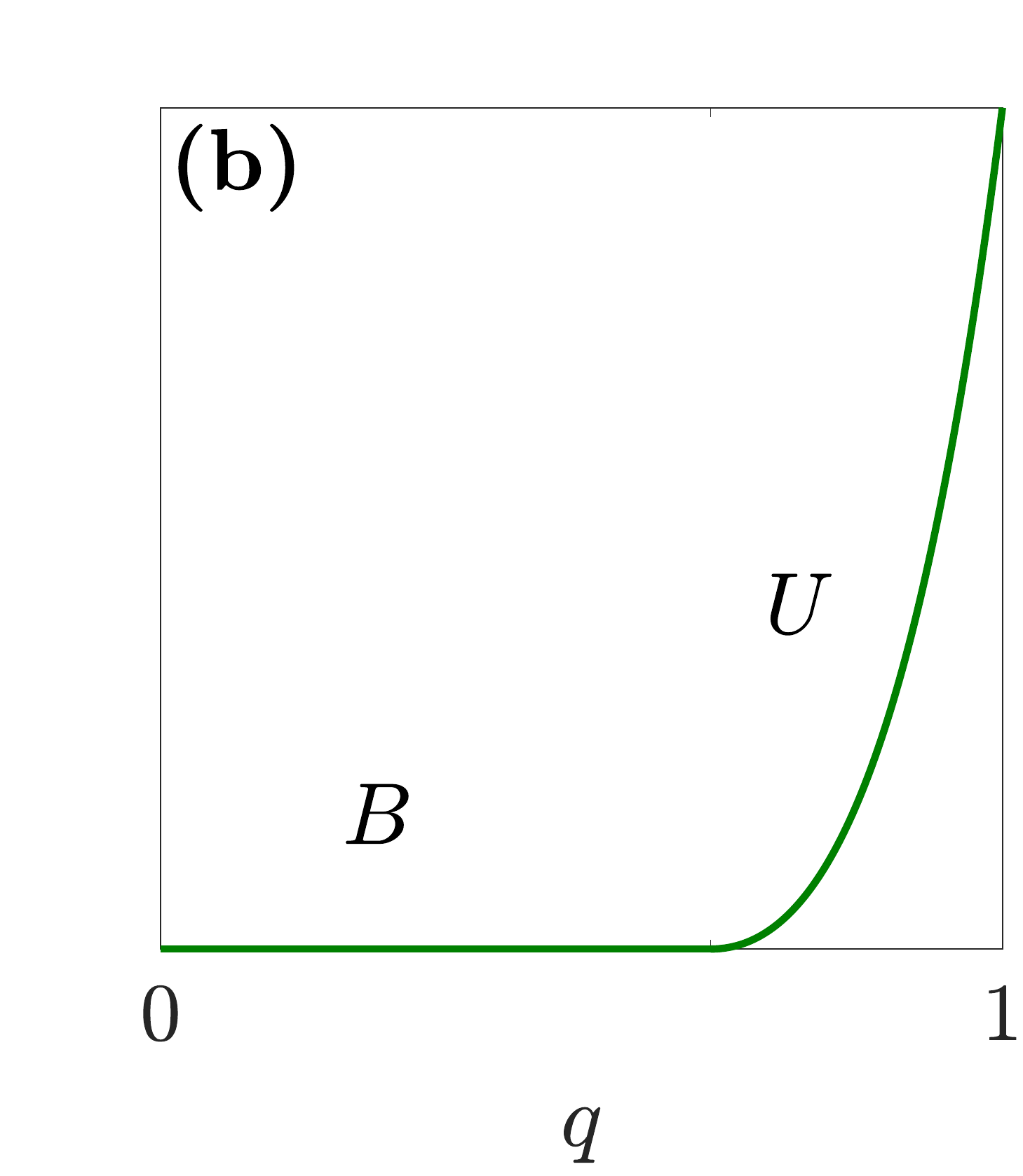}
	\end{center}
	\caption{\label{fig:rcurve_kgt0_g1} The shape of the longest curve for random boarding can be either ordinary or piecewise. 
		(a) Ordinary-type curve ($O$) when $0<k\leqslant\ln(2)$. 
		(b) Piecewise curve consisting of a constant function along the base ($B$) which is smoothly continued by an upward-going ordinary-type curve ($U$) when $k>\ln(2)$.}
\end{figure*}

The constants $a,b$ are determined using the values at the start and end points: $r^*(0)=0$ and $r^*(1)=1$. 
A typical shape is shown in \crefformat{figure}{Fig.~#2#1{(a)}#3}\cref{fig:rcurve_kgt0_g1} for $k\leqslant \ln(2)$. 

However, when $k> \ln(2)$, $r^*(q)$ will extend below the $(q,r)$-unit square. Since the curve should be within the unit square, a piecewise curve as in \crefformat{figure}{Fig.~#2#1{(b)}#3}\cref{fig:rcurve_kgt0_g1} emerges as the longest curve (see \cite{Erland/Kaupuzs/Frette/Pugatch/Bachmat:2019} for details).
 
The lengths of these longest curves are computed by \cref{eq:length}, and the expected boarding time with random boarding and effective aisle-clearing time $\tau_A$ is by leading order given by \cref{eq:Tapprox}, such that
\begin{equation}\label{eq:T_R}
\hat{T}_\textrm{RA} =
\begin{cases}
2 \tau_A\sqrt{\frac{ N}{k}}\sqrt{e^k-1}
\qquad &  0<k\leqslant \ln(2) \\[0ex]
2 \tau_A\sqrt{\frac{ N}{k}}\left[k-\ln(2)+1\right]
\qquad & \ln(2) < k.
\end{cases}
\end{equation}

\begin{figure*}[t!]
	\begin{center}
		\includegraphics[width=8cm,clip]{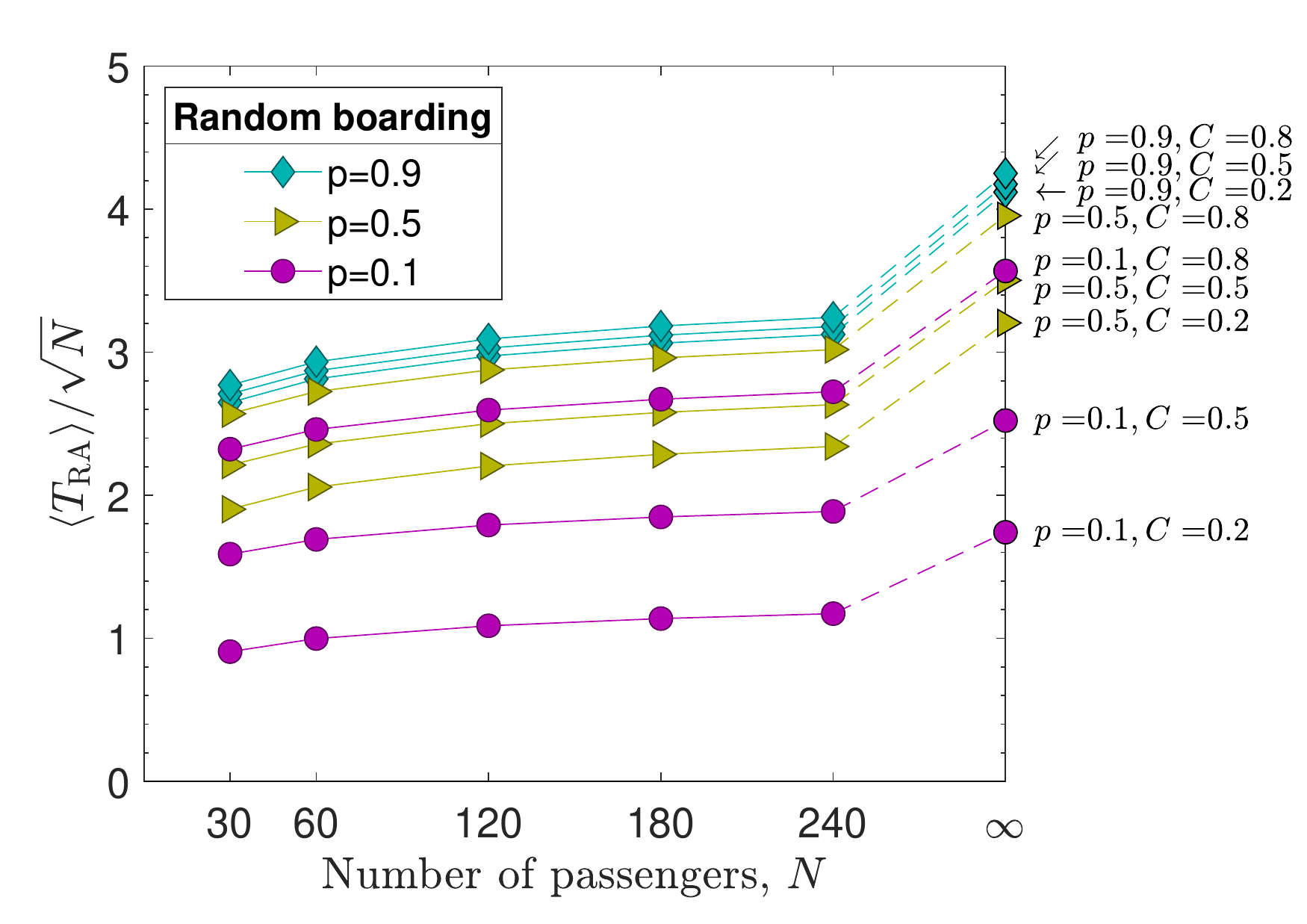}
	\end{center}
	\caption{\label{fig:RA_boardingtimes_Ndevelop} Average boarding time estimates for the random boarding  policy for different $(p,C)$-parameter settings. Simulation results for increasing number of passengers are compared to the asymptotic results for all combinations of parameter values $p\in\{0.1,0.5,0.9\}$ and $C\in\{0.2,0.5,0.8\}$. $k=4$, $\tau_S=1$, and the accuracy is $\pm 0.002$ (as a result of $10^6$ {runs} for each finite-$N$ data point). The rightmost points are the asymptotic values in \cref{eq:T_R}. Corresponding results are presented in \cite{Erland/Kaupuzs/Frette/Pugatch/Bachmat:2019} for the slow-first and fast-first policies. 
	}
\end{figure*}

In Fig.~\ref{fig:RA_boardingtimes_Ndevelop} comparisons of the asymptotic boarding time for the random boarding policy with simulation results for $N\leqslant 240$ show that the asymptotic result in \cref{eq:T_R} tends to overestimate the boarding time. Still, the relative ranking between different parameter settings is maintained. 
$\tau_A$ is here estimated by extensive discrete-event simulations to a high degree of accuracy as in \cref{ssec:jevgenijs}. 

\subsection{Analysis of the slow-first policy}
When there are two groups, as in, e.g., the slow-first policy, the curve weight for a function $r(q)$ on the interval $q\in (0,1)$ is given by
\begin{align} 
W_\textrm{SF}(r) &= \int_{0}^{p} \tau_S \sqrt{r'(q) + k(1-r(q))}dq\nonumber\\ 
&\quad + \int_{p}^{1} \tau_F \sqrt{r'(q) + k(1-r(q))}dq \nonumber\\[1ex]
&= \tau_S L_S(r) + \tau_F L_F(r) \label{eq:weight_SF}
\end{align}
where $L_S,L_F$ are curve lengths as defined in \cref{eq:length} and $\tau_S,\tau_F$ are the effective aisle clearing time for the slow and fast group, respectively. 

\begin{figure*}[t!]
	\begin{center}
		\includegraphics[width=16.0cm,clip]{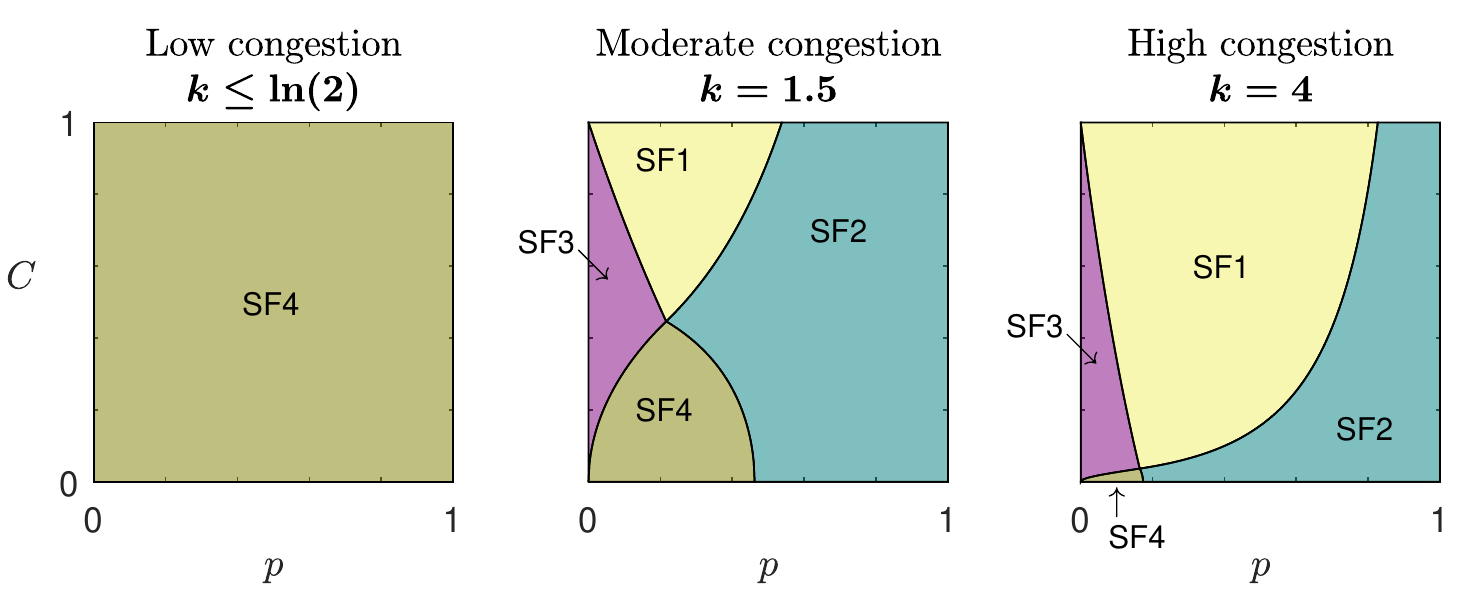}
	\end{center}
	\caption{\label{fig:SF_activesubfunctions} The subdomains of the $(p,C)$-unit square where the slow-first boarding time {and the maximum of $W_\textrm{SF}$ in \cref{eq:weight_SF}} is represented by the different sub-functions in \cref{eq:W*_SF}.
	}
\end{figure*}

The maximal weight $W_\textrm{SF}^*$ of $W_\textrm{SF}$ is given by the curve that gives the maximal curve weight in \cref{eq:weight_SF}. The shape of the heaviest curve is determined by the values of $k,p,C$ where $C\equiv \tau_F/\tau_S \in(0,1)$. Detailed computations of the heaviest curves and their respective curve weights are presented in \cite{Erland/Kaupuzs/Frette/Pugatch/Bachmat:2019}. The expression for the maximal curve weight $W_\textrm{SF}^*(k,p,C)$ is also given in \cref{app:proofs} and consists of four different subfunctions. Fig. \ref{fig:SF_activesubfunctions} illustrates how the subdomains of each of the subfunctions depend on the parameters. 
For example, for $k \leqslant \ln(2)$, only one of the subfunctions (SF4) are present, whereas another subfunction (SF1) dominates the parameter space when $k\rightarrow\infty$.

The maximal weight $W_\textrm{SF}^*$ is used to calculate the corresponding asymptotic boarding time $\hat{T}_\textrm{SF}$ in \cref{eq:Tapprox}. Comparisons of the asymptotic boarding time for the slow-first policy with simulation results for $N\leqslant 240$ in \cite{Erland/Kaupuzs/Frette/Pugatch/Bachmat:2019} show that the asymptotic result in \cref{eq:Tapprox} tends to overestimate the boarding time, but the relative ranking between different parameter settings is maintained.
Corresponding analytical expressions and results for the fast-first policy are also presented in \cite{Erland/Kaupuzs/Frette/Pugatch/Bachmat:2019}.

\subsection{Comparison of slow-first and random boarding policies}\label{ss:RAvsSF}
As stated in \cref{th:RAvsSF} in \cref{sec:mainresults}, the slow-first policy outperforms the random boarding policy for all values of $k>0$ and $p,C \in (0,1)$. The results are valid for concave mixtures and the proof showing that $W^*_\textrm{RA}-W^*_\textrm{SF}>0$,  is left for Appendix \ref{app:proofs}. The results of the finite-$N$ discrete-event simulations in \cref{fig:mainresults,fig:RAvsSFcontourplot} indicate that the result is also valid for realistic numbers of passengers $N$.

\begin{figure*}[htb]
	\begin{center}
		\includegraphics[width=8cm,clip]{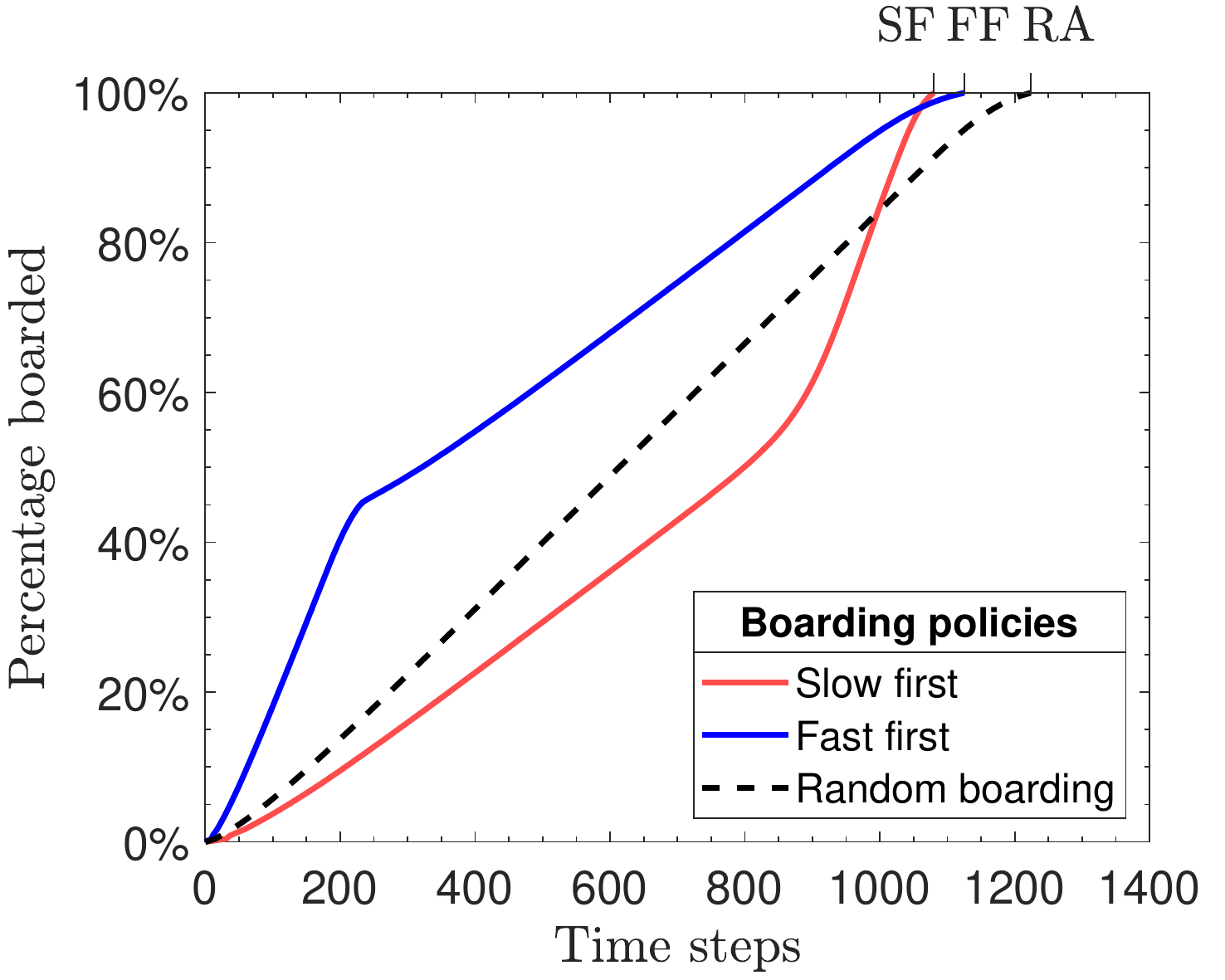}  
	\end{center}
	\caption{\label{fig:timedevelop} 
		Comparison of 3 different boarding policies. We used parameters based on the empirical data in \cref{ssec:empirical}, with congestion $k=4$, $55\%$ slow passengers ($p=0.55$) and assumed that the fast passengers (those without hand luggage) clear the aisle three times faster than the slow passengers ($C=0.3$). We assumed there are 6 seats per row, and a total of $N = 240$ passengers. The percentage of seated passengers is plotted as a function of time. 
		On average the slow-first policy (rightmost, red, solid line) is lagging behind all the way to around $\sim 80\%$. However, the slow-first policy eventually seats all passengers in a shorter time --- relative to the two other policies.
		Fast first is second (FF; $+4\%$), and the random boarding policy turns out to be the worst (RA; $+13\%$).
		The graph is an average of 10,000 discrete-event runs.
	}
\end{figure*}

Another feature of the result is presented in Fig.~\ref{fig:timedevelop} where the percentage of seated passengers is plotted as a function of time for the three different policies. The same empirical parameter setting as in Fig.~\ref{fig:policies} are used.
The boarding time $T$ is equal to the time when the fraction of seated passengers reaches $100\%$. Slow first ranks first ($T_\textrm{SF}=1080$ time steps), then fast first ($T_\textrm{FF}=1125$) and at last random boarding ($T_\textrm{RA}=1224$).  

The graph  for the slow-first policy (SF) in Fig.~\ref{fig:timedevelop} consists of two curve segments with different slopes. The first, less steep segment corresponds to boarding dominated by slow passengers, while the steep segment is dominated by fast passengers. As boarding starts, the queue of passengers is four times as long as the aisle ($k=4$), and the first fast passengers in slow first enter the airplane only after a significant portion of the slow passengers is seated.

The contour plot in \crefformat{figure}{Fig.~#2#1{(a)}#3}\cref{fig:RAvsSFcontourplot} indicates that the relative distance between random boarding and slow first increases for decreasing $C$. This is explicitly shown in \cref{eq:RArelSF1partialC,eq:RArelSF4squaredpartialC2} for the SF1 and the SF4 regions, respectively. Ultimately, this leads to the maximum relative distance being obtained for small $C$, as stated in \cref{eq:RAvsFFmaxmain}. 


As indicated in Fig.~\ref{fig:SF_activesubfunctions}, $W^*_\textrm{SF}=W^*_\textrm{SF1}$ for large $k$ when $p,C$ are fixed. In \cref{app:RAvsSF_kinfty} we show that
\begin{align}\label{eq:RArelSFkinfty}
\frac{W^*_\textrm{RA}}{W^*_\textrm{SF}}
&\overset{k\rightarrow \infty}{\longrightarrow} \quad 
\frac{\sqrt{p+C^2(1-p)}}{p+C(1-p)} \quad >1.
\end{align}
This explains why the relative difference between slow first and random boarding in Fig.~\ref{fig:mainresults} approaches a limiting value when $N,k\rightarrow\infty$. That both the slow-first and the fast-first policies approach the same limiting values is shown in \cite{Erland/Kaupuzs/Frette/Pugatch/Bachmat:2019}. The limiting value in \cref{eq:RArelSFkinfty} increases for smaller $C$, and for fixed $C$, \cref{eq:RArelSFkinfty} is maximized by $p=C/(1-C)$, giving the maximum $(1+C)/(2\sqrt{C})$.

\subsection{Comparison of fast-first and random boarding policies}
\begin{figure*}[h!]
	\begin{center}
		\includegraphics[width=8cm]{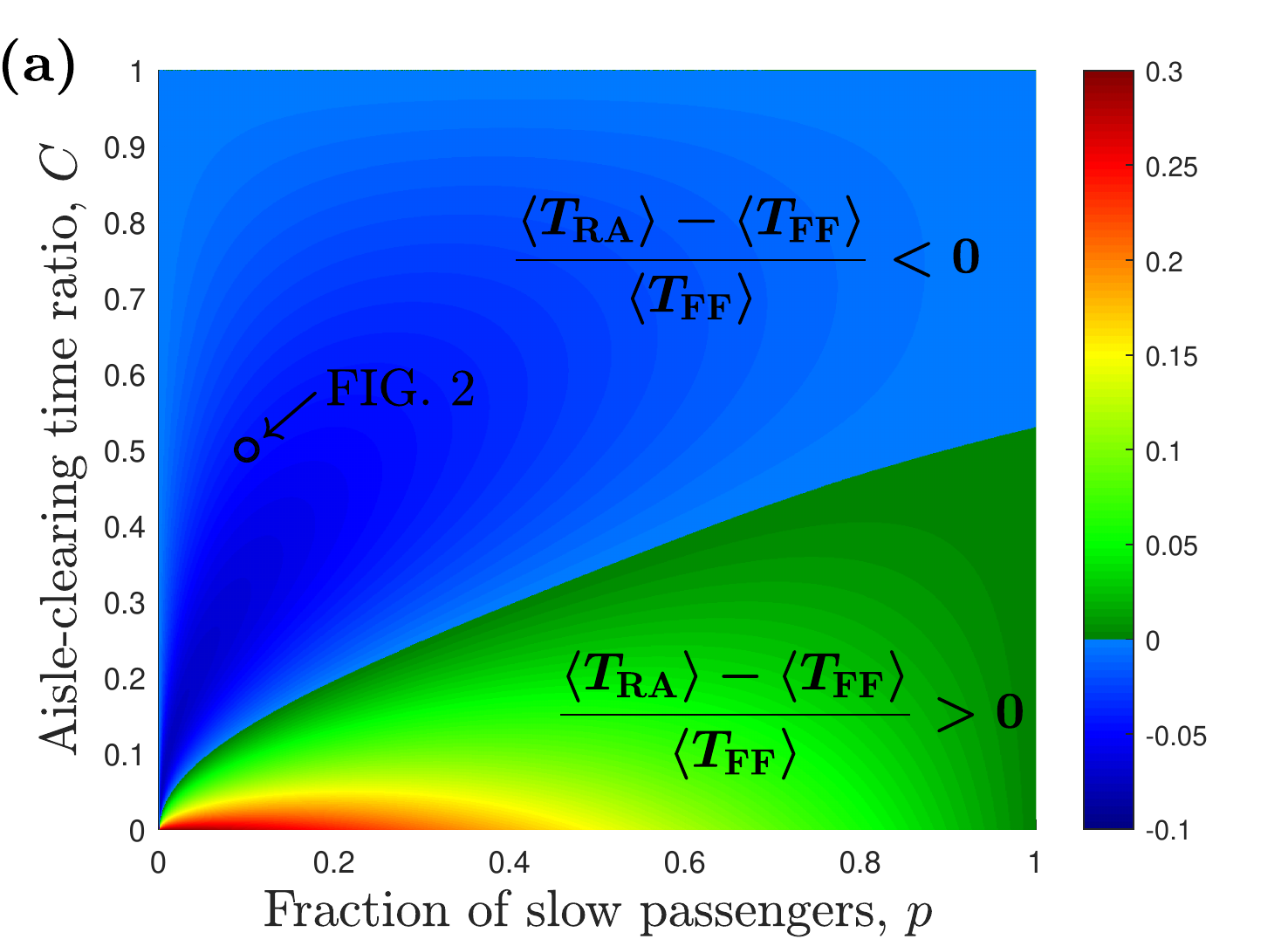}
		\parbox[b][3.6cm][c]{0.1cm}{$\quad$}
		\includegraphics[width=8cm]{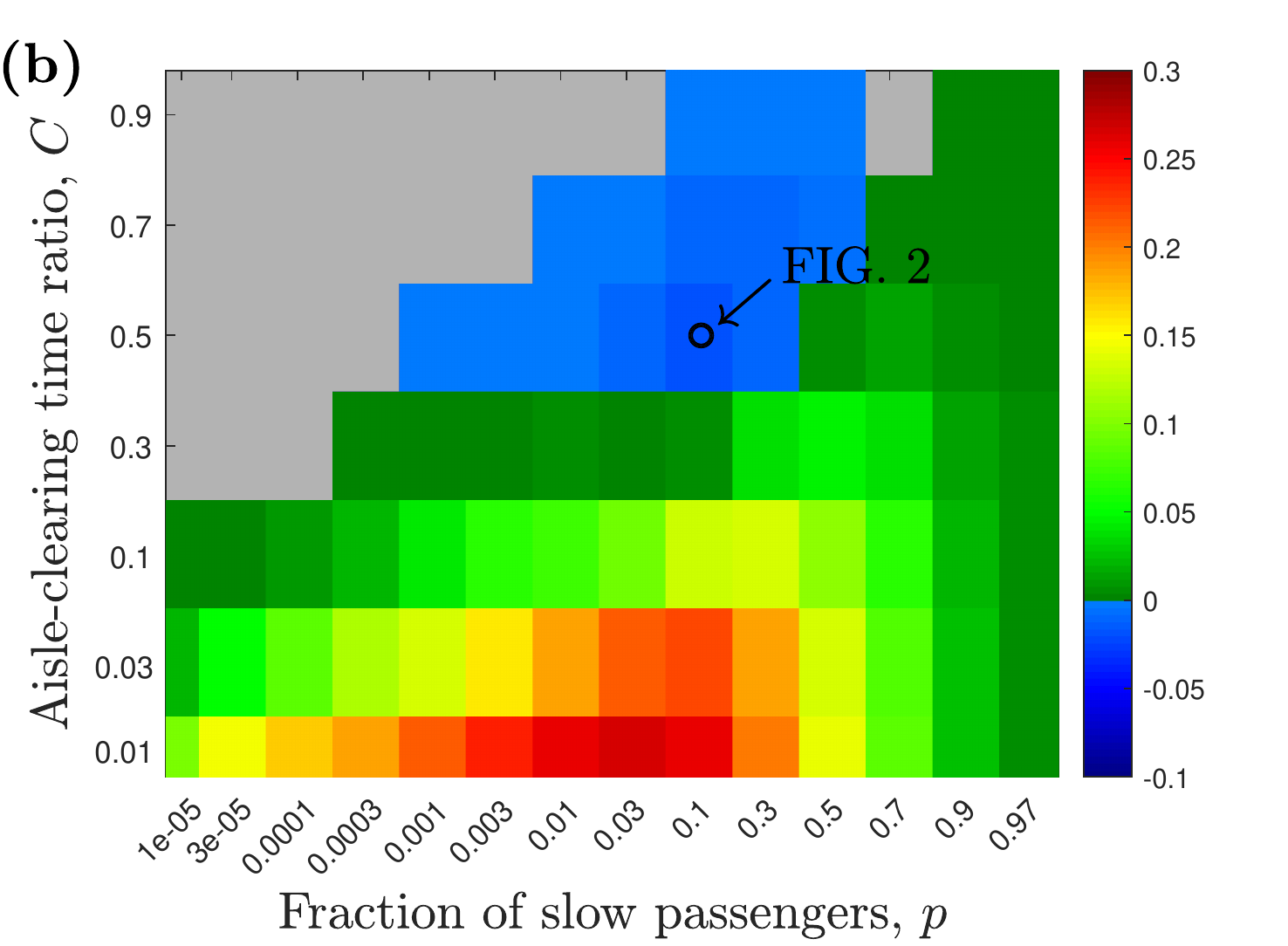}
	\end{center}
	\caption{\label{fig:RAvsFFcontourplot} Relative difference in asymptotic average boarding time $(\angles{T_\textrm{RA}}-\angles{T_\textrm{FF}})/\angles{ T_\textrm{FF}}$ between the random boarding and the fast-first policies when $k=1$ {as in Fig.~\ref{fig:mainresults}}. (a) The effective aisle-clearing time of random boarding $\tau_A$ is approximated by $\hat{\tau}^2_{A,mix}\equiv p\tau_S^2+(1-p)\tau_F^2$, and fast first is seemingly inferior to random boarding on large portions of the $(p,C)$ parameter space. (b) The precise estimates of $\tau_A$ in \cref{ssec:jevgenijs} are applied, hence the asymptotic total boarding time of random boarding is larger than in (a). Still, fast first is inferior to random boarding for several values of $p$ small and $C$ large. Gray areas indicate that the ranking of the two policies was indeterminate within the range of $\pm 2*$std.dev of the $\tau_A$ estimate.
	}
\end{figure*}

As opposed to the slow-first policy, fast first can be both better and worse than random boarding, depending on the congestion $k$ as shown in Fig.~\ref{fig:mainresults} for $p=0.1, C=0.5$. The same feature is shown for other $(p,C)$-values in Fig.~\ref{fig:RAvsFFcontourplot} when $k=1$ is fixed. The {blue-shaded} areas in the figure indicate that the asymptotic boarding time of random boarding is shorter than for the fast-first policy (negative relative difference).

The effective aisle-clearing time of random boarding $\tau_A$ is approximated by $\hat{\tau}^2_{A,mix}$ in the estimates in \crefformat{figure}{Fig.~#2#1{(a)}#3}\cref{fig:RAvsFFcontourplot} and the results indicate that fast first can be inferior to random boarding for $N=\infty$. In \crefformat{figure}{Fig.~#2#1{(b)}#3}\cref{fig:RAvsFFcontourplot} this is verified for several combinations of $p,C$-values when precise estimates of $\tau_A$ are applied. 

\begin{figure*}[h!]
	\begin{center}
		\includegraphics[width=8cm]{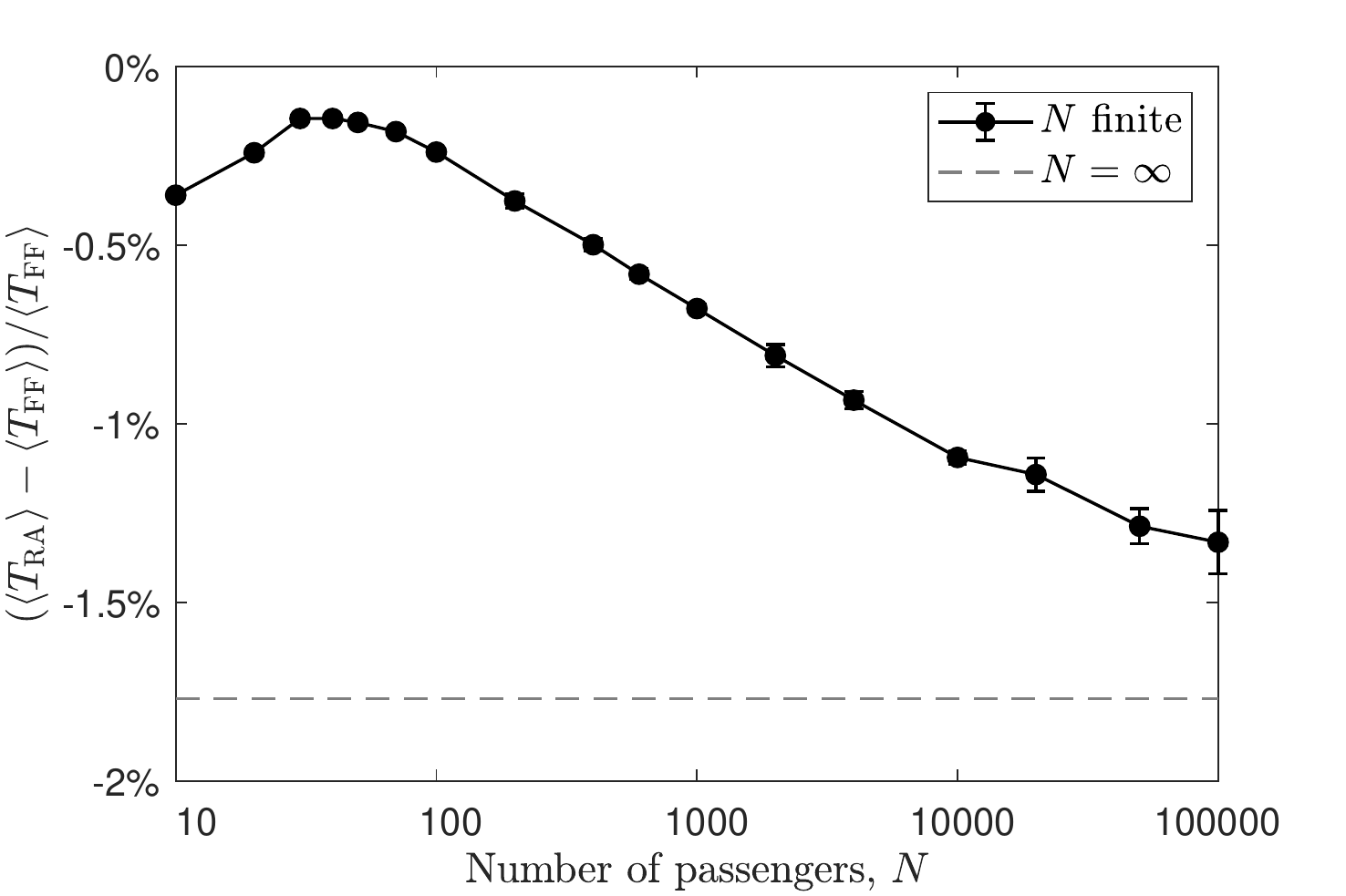}
	\end{center}
	\caption{\label{fig:RAvsFF_Ndevelop} 
	Relative difference in asymptotic average boarding time 
	$(\angles{T_\textrm{RA}}-\angles{T_\textrm{FF}})/\angles{ T_\textrm{FF}}$ between the random boarding and the fast-first policies when $(k,p,C)=(1,0.1,0.5)$. For this setting the graph indicates that random boarding is superior also for finite $N$. For large $N$ the relative difference approaches -1.77\% which is the asymptotic value ($N\rightarrow\infty$) marked with a black circle in \cref{fig:RAvsFFcontourplot}. 
	}
\end{figure*}

Fig.~\ref{fig:RAvsFF_Ndevelop} demonstrates that fast first is inferior to random boarding also for finite number of passengers $N$ for the particular set of parameters $k=1,p=0.1,C=0.5$ (black circles in Fig.~\ref{fig:RAvsFFcontourplot}). For large $N$ the relative difference approaches $-1.77\%$ which corresponds to the minimum value in \crefformat{figure}{Fig.~#2#1{(b)}#3}\cref{fig:RAvsFFcontourplot} (close to the peak value of the blue dashed curve in Fig.~\ref{fig:mainresults}).


\subsection{Optimal separation strategies based on empirical data}\label{ssec:Steinerdatacomparisons}
\begin{figure*}[htb]
	\begin{center}
		\includegraphics[width=8cm]{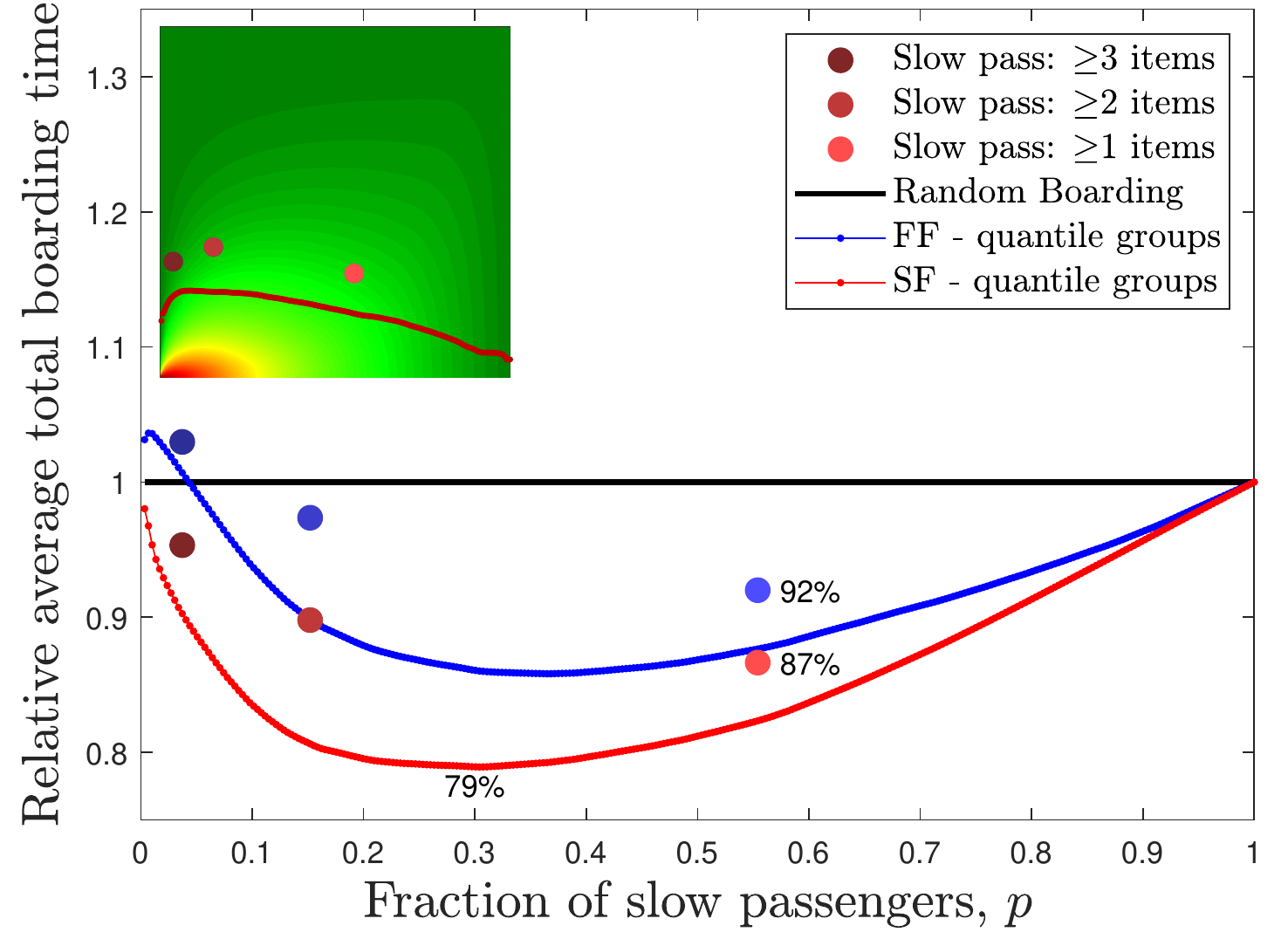}
	\end{center}
	\caption{\label{fig:AsymptoticSteinerdata}  		
		Relative asymptotic boarding time for slow first (SF, {red}) and fast first (FF, {blue}) compared to random boarding for different ($p,C$)-values based on empirical data. Results for luggage-based  separation strategies are shown as red- and blue-colored bullets, while separations that are {based on quantiles of the empirical aisle-clearing time distribution} are shown as dot-lined curves.
		Red-colored results are asymptotic values {taken from} the indicated positions in the inset contour plot of the relative difference between random boarding and slow first from \cref{fig:RAvsSFcontourplot}. The effective aisle-clearing time of random boarding $\tau_A$ is approximated by $\hat{\tau}^2_{A,mix}\equiv p\tau_S^2+(1-p)\tau_F^2$ ({conjectured as} a lower bound).
		{If the slow group are those with 1 luggage item or more, the improvement by choosing slow first instead of random boarding is 13\%. The corresponding result for fast first  is 8\%.} However, separation where slow passengers have three or more luggage items results in a fast-first policy which could be inferior to random boarding. The improvement is {up to} 21\% when separating slow and fast groups according to quantiles in the aisle-clearing time distribution. 
	}
\end{figure*}

The asymptotic boarding times for random boarding, slow first (SF) and fast first (FF) based on the ($p,C$)-values for the different separation strategies in \cref{ssec:empirical} are compared in Fig.~\ref{fig:AsymptoticSteinerdata}. The congestion parameter is set to $k=4$ (typical empirical value) {and we assume that $\tau_A = \hat{\tau}^2_{A,mix}$}. The inset from \crefformat{figure}{Fig.~#2#1{(a)}#3}\cref{fig:RAvsSFcontourplot} shows that the ($p,C$)-values (bullets) are in the vicinity of the region with largest improvement {for} the slow-first policy relative to random boarding. The asymptotic boarding time of slow first and fast first relative to random boarding is shown as red and blue bullets, respectively.

The improvement by choosing the slow-first policy (and for most parameter settings also fast first) instead of random boarding is significant. From Fig.~\ref{fig:AsymptoticSteinerdata} one sees that when the slow and fast groups are separated based on the number of items, the maximum improvement with slow first compared to random boarding is {obtained when the slow group are those who carry luggage items. Application of precise estimates of $\tau_A$ from \cref{ssec:empirical} (instead of assuming $\tau_A = \hat{\tau}^2_{A,mix}$), gives a slight adjustment from $13\%$ to $14\%$ improvement. For finite number of passengers $N=240$, the improvement is $13\%$ for the same parameter settings (see \cref{fig:timedevelop}).} 

When the slow and fast groups are separated (clairvoyantly) based on those $p=30\%$ who have the slowest aisle-clearing times, the maximum improvement with slow first compared to random boarding is 21\% {(when assuming $\tau_A=\hat{\tau}^2_{A,mix}$)}.  The choice of $p$ is quite robust to variations as the improvement is quite similar for all values of $p$ in the range $p\in[0.2,0.4]$.


\clearpage
\section{Discussion and outlook}
In this paper we recast the airplane boarding problem to the setting of Lorentzian geometry and show that the slow-first boarding policy is superior to random boarding for any set of parameters when the number of passengers $N\rightarrow\infty$. 
The analytical result is presented in \cref{th:RAvsSF} and hinges on a concavity assumption regarding the effective aisle-clearing time of groups of passengers with varying aisle-clearing times. This parameter is not available in closed form, but extensive numerical computations enable us to validate the concavity assumption with a high degree of certainty.

As stated in \cref{eq:RAvsFFmaxmain}, the maximum relative distance between random boarding and slow first is obtained for a small ratio $C$ between effective aisle-clearing times of the fast and the slow group.
This seems to be an important feature that distinguishes the slow-first policy from random boarding.
$C$ being small implies that the aisle-clearing time of the slow passengers is much longer than for the fast passengers. In random boarding a few slow passengers will then be able to block and delay many fast passengers from getting to their seats, in particular when the congestion $k$ is large. 
With the slow-first policy the variability within each group is reduced, and this enables more synchronized clearing of the aisle and less effect of local interference between passengers. Since the relative difference tend to increase with $C$ small and $k$ large, local interference seems to be the crucial detrimental aspect of random boarding. This aspect seems to be much less pronounced within the more homogeneous groups in the slow-first policy.

While slow first is always superior to random boarding, the ranking of fast first and random boarding depends on the values of the parameters $(k,p,C)$.  Fast first enables synchronized seating in the same way as slow first, and the positive effect of this is higher when $C$ is small, and $k$ is large. However, random boarding is superior for large $C$ and small $k$.  A heuristic explanation for the non-consistent ranking of the two policies could be found in Fig.~\ref{fig:timedevelop}. The graphs indicate that the rate of seating is slower in the start phase for all boarding policies. While the slow-first policy enables parallel seating in the transition phase between the groups \cite{Erland/Kaupuzs/Frette/Pugatch/Bachmat:2019}, there is in effect two starting phases with the fast-first policy --- one for the fast and one for the slow group. For certain parameter settings (e.g. large $C$ and small $p$), the positive effect of improved local synchronization in fast first is not enough to offset this.


The present paper together with \cite{Erland/Kaupuzs/Frette/Pugatch/Bachmat:2019} provides a complete picture of the relation between three policies which are actually practiced by airlines --- random boarding, fast first and slow first across all parameter settings. In ongoing work with several additional co-authors, we are extending this work in new directions, both in terms of new optimization functions and new policies, in terms of finding optimal policies, and considering more than two groups. 

Considering more than two groups, a possible extension of the slow first policy could be to distribute the passengers into more groups with even less variation within each group. Since slow first is better, the groups should be sequenced in the queue according to descending effective aisle-clearing times. In fact, preliminary results indicate that this strategy reduces the total boarding time, and ultimately, a queue where all passengers are ranked according to descending aisle-clearing time is even better. 

With respect to optimization function, one could consider average individual boarding time (customer experience) rather than total boarding time. Then the ranking of the three policies slow first, random boarding and fast first is reversed, and the performance differences between fast first and slow first are more pronounced compared to total boarding time. Consequently, the total boarding time of slow first is shorter, but the average passenger suffers longer waits in the queue. This leads to the introduction of new policies which interpolate between slow first and fast first and provide the benefits of both. 

The problem of finding optimal queue-row placements is also of great interest. Such placements form space-time lenses with respect to proper time for positive mass particles, in the sense that $(0,0)$ and $(1,1)$ become conjugate points. This leads to some efficient lens constructions, but as boarding policies they are often very difficult to implement. However, for large values of $k$, there are near optimal lenses with small amounts of aberration to make the boarding rules reasonable.

\begin{acknowledgments}
	The work of Eitan Bachmat was supported by the German Science Foundation (DFG) through
	the grant “Airplane Boarding” (JA 2311/3-1).
	The authors acknowledge the use of resources provided by the Latvian Grid Infrastructure and High Performance Computing centre of Riga Technical University.	
\end{acknowledgments}

\clearpage
\appendix
\section{Procedure for estimation of $\tau_X$}\label{app:HIS}

In this appendix we describe an efficient procedure for the estimation of the effective aisle-clearing time $\tau_X$, when the aisle clearing time $X$ of all passengers in a group follow a common distribution. Since $\tau_X$ is independent of $k$, we refer to the case when $k=0$. 
	
First, we establish an efficient algorithm for the computation of the boarding time $T$ for a given queue, by recasting the problem to the computation of a heaviest increasing subsequence (HIS). We use a variant of the algorithm in \cite{Jacobson/Vo:1992}. Then $\angles{T}$ is estimated by simulation for a series of $N$ values. The first order approximation $\angles{T} / (2 \sqrt{N}) \rightarrow \tau_X$ (see \cref{sec:analysis}) is improved by an extrapolation to $N \to \infty$ using the extended large $N$ behavior in \cref{eq:tau_asymptotics}.

We first considered an example, where the weights $\tau_F=1$ (fast passengers) and $\tau_S=2$ (slow passengers)
are assigned randomly, each with probability $p=1/2$ (the fraction of slow passengers).
We tested the algorithm for $N=1000, 8000, 64 \, 000$ and $512 \, 000$ passengers. $10 \, 000$ simulation runs was
performed for $N \le 8000$ and $1000$ simulation runs -- for $N \ge 64 \, 000$. 

The estimated average boarding times $\langle T \rangle$ for random boarding (the values of HIS), normalized to $\sqrt{N}$ are given in \cref{tab1}. 
The asymptotic value at $N=\infty$ is obtained by an extrapolation (see the following section), taking into account that corrections to scaling can be expanded in powers of $N^{-1/3}$ as in \cref{eq:tau_asymptotics}.

\begin{table}[h]
	\caption{The dependence on $N$ of the normalized boarding time $\langle T \rangle/\sqrt{N}$ of random boarding when $p=0.5=C$.}
	\label{tab1}
	\begin{center}
		\begin{tabular}{|c|c|}
			\hline
			\rule[-2mm]{0mm}{7mm}
			N & $\langle T \rangle/\sqrt{N}$  \\
			\hline
			1000 & 2.9802(14)  \\
			8000 & 3.11190(75)  \\
			64000 & 3.1798(12)  \\
			512000 & 3.21753(61) \\
			$\infty$ & 3.2553(17) \\
			\hline
		\end{tabular}
	\end{center}
\end{table}

The normalized asymptotic boarding time for both slow-first and fast-first is just $2 \sqrt{\langle X^2 \rangle} \approx 3.1622777$ (when $k=0, p=0.5, C=0.5$). It is seen that, in this particular case, the random boarding policy is a bit worse (slower).

\subsection{$\tau_X$ for two-valued distributions}
In order to estimate the unknown constant $\tau_X$, and to investigate whether it is bounded from below by $\sqrt{\langle X^2 \rangle}$, we take the asymptotic result in \cref{eq:tau_asymptotics} and define the ratio 
	\begin{equation}\label{eq:phi}
		\phi_N \equiv \frac{{T}/\sqrt{N}}{2\sqrt{\langle X^2 \rangle}} \;\sim\; \phi_\infty - \tilde{a}_1 N^{-\frac{1}{3}}.
	\end{equation}
	where $\phi_\infty \equiv \tau_X/\sqrt{\langle X^2 \rangle}$. Then the following extrapolation algorithm to estimate $\phi_\infty$ has been used.

We performed simulations at a hierarchy of $N=N_i$ values: $N_0$, $N_1=8N_0$, $N_2=8^2N_0$, etc.. This resulted in a corresponding hierarchy of $\phi_N$ values, i.~e., ${\phi}_{N_0}$, ${\phi}_{N_1}$, ${\phi}_{N_2}$, etc. Using that $\phi_N$ in \cref{eq:phi} is linear in $N^{-1/3}$, we obtain a sequence of estimates for $\phi_\infty$ by linear extrapolation, i.~e.,
\begin{equation}\label{eq:phi_linear}
	{\phi}^{\mathrm{lin}}_i = 2{\phi}_{N_i} - {\phi}_{N_{i-1}}
\end{equation}
for $i \ge 1$. 
With the ansatz that \cref{eq:phi} can be expanded further in powers of $N^{-\frac{1}{3}}$, we also considered the quadratic extrapolation:
\begin{equation}\label{eq:phi_quad}
	{\phi}^{\mathrm{quad}}_i = \displaystyle{\frac{1}{3}} \left(8{\phi}_{N_i} - 6{\phi}_{N_{i-1}} + {\phi}_{N_{i-2}} \right).
\end{equation}

Each next value is expected to be twice closer to the asymptotic value at $N \to \infty$ for the original sequence ${\phi}_{N_i}$, 4 times closer for ${\phi}^{\mathrm{lin}}_i$ and 8 times closer for ${\phi}^{\mathrm{quad}}_i$. 
We also observed that, at a given number $M$ of simulation runs, each next value in any of these sequences has approximately twice smaller statistical error (standard deviation $\sigma$).

We used the following criterion: the maximal number of points (passengers) $N$ has to be reached, which is large enough to ensure that the systematical extrapolation error is much smaller than $\sigma$. The statistical error
for ${\phi}^{\mathrm{lin}}_i$ is somewhat smaller than that of ${\phi}^{\mathrm{quad}}_i$, however, in some of the considered cases the above criterion was not well satisfied for ${\phi}^{\mathrm{lin}}_i$, but it was satisfied for ${\phi}^{\mathrm{quad}}_i$. Therefore, ${\phi}^{\mathrm{quad}}_i$ values for the largest $i$ 
have been used as the final asymptotic estimates of .

Calculations within the range of parameters $0.01 \le C \le 0.9$ and $0.00001 \le p \le 0.97$ have been
performed. The simulation parameter $N_0$ ranged from 10 to 1000 (larger for smaller $C$ and $p$), whereas maximal $i$ has been fixed equal to 7. 
The number of MC realizations (simulation runs) ranged from $10\,000$ (for $N_0=10$) to $500$ (for $N_0=1000$). The results are
collected in \cref{tab:tau_twovalued}.

The most accurate value has been obtained for $p=0.97$ and $C=0.01$. In this case, the original
less accurate value $0.99981(16)$ was by $\approx 1.2 \sigma$ smaller than unity, therefore
we performed extra simulations to verify whether or not it is $<1$. As a result, the possibility that
$\phi_\infty<1$ holds in this case has not been confirmed. In all other cases $\phi_\infty$ is either $>1$, or the deviation below unity is within one $\sigma$. 

Thus, our simulation and calculation results do not allow us to conclude
that $\tau_X$ in some cases is smaller than $\sqrt{\langle X^2 \rangle}$ for the two-value distribution.

\subsection{$\tau_X$ for multi-valued distributions based on empirical data}

We consider the aisle clearing time distribution data in \cite{Steiner:2008}.
The mean boarding time $\langle T \rangle$ for the multi-value distribution is determined by the HIS algorithm where each weight is taken from the same distribution. 
For each distribution we define a hierarchy of $\tau_{X,N}$ values where for each $N$, $\tau_{X,N} = \angles{T}/ (2 \sqrt{N})$. As for the two-valued distributions, we use linear and quadratic polynomials of $N^{-1/3}$ (corresponding to \cref{eq:phi_linear,eq:phi_quad}) for an extrapolation to $N=\infty$, based on the idea that $\angles{T}/\sqrt{N}$ can be expanded in powers of $N^{-1/3}$. The results for $10000$ simulation runs are collected in \cref{tabula} for the separation strategy when the slow passengers are those carrying hand luggage.

\begin{table}[h]
	\caption{The values of $\tau_{X,N}$ (${\tau_A}$ for all passengers, $\tau_F$ for fast passengers and $\tau_S$ for slow passengers when the slow passengers are those carrying hand luggage),
		extracted from $10000$ runs of HIS simulations depending on the number of passengers $N$. The asymptotic estimates are obtained by a linear and a quadratic extrapolation in the variable $N^{-1/3}$.}
	\label{tabula}
	\begin{center}
		\begin{tabular}{|c|c|c|c|}
			\hline
			\rule[-2mm]{0mm}{7mm}
			$N$ & ${\tau_A}$ & $\tau_F$ & $\tau_S$   \\
			\hline
			100 & 0.33337(62)  & 0.12969(22) & 0.44204(70) \\
			800 & 0.38168(35)  & 0.14771(12) & 0.49844(39) \\
			6400 & 0.41048(18)  & 0.158080(65) & 0.53049(20) \\
			51200 & 0.426061(93) & 0.163696(33) & 0.54820(10) \\
			409600 & 0.434405(48) & 0.166659(17) & 0.557788(53) \\
			3276800 & 0.438745(24) & 0.1682272(86) & 0.562650(27) \\
			\hline
			$\infty$ (linear extr.) & 0.443086(68) & 0.169795(24) & 0.567511(75) \\
			\hline
			$\infty$ (quadr. extr.) & 0.44320(12)  & 0.169853(43) & 0.56756(13) \\
			\hline
		\end{tabular}
	\end{center}
\end{table}

A fast convergence of a series of extrapolated values shows that such an extrapolation is accurate enough. 
The values of the linear extrapolation in \cref{tabula} have smaller statistical errors, but systematical extrapolation errors 
are smaller and practically negligible for the values of the quadratic extrapolation. 
Hence ${\tau_A}=0.44320(12)$, $\tau_F=0.169853(43)$ and $\tau_S=0.56756(13)$ can be assumed as the final estimates. The estimates for $\tau_F$ and $\tau_S$ in \cref{tab:tau_Steiner} when the passengers are divided into fast and slow groups in other ways are obtained in a similar manner.

\section{Proofs}\label{app:proofs}
There are five different combinations of RA and SF subfunctions which we treat in the subsections below in order to show that $W^*_\textrm{RA}-W^*_\textrm{SF}>0$ (for $k\leqslant\ln(2)$ there is one combination, and for
$k>\ln(2)$ there are four combinations).

\begin{equation*}
W_\textrm{RA}^* =
\begin{cases}
\tau_{A} \sqrt{\frac{1}{k}}\sqrt{e^k-1}
\qquad &  0<k\leqslant \ln(2) \\[0ex]
\tau_{A} \sqrt{\frac{1}{k}}\left[k-\ln(2)+1\right]
\qquad & \ln(2) < k.
\end{cases}
\end{equation*}
We assume that
\begin{equation*}
	\tau_{A}^2 
	= p\tau_S^2 + (1-p)\tau_F^2  
	= \tau_S^2 \left[p + C^2(1-p)\right]
	\equiv \hat{\tau}^2_{A,mix},
\end{equation*}
and prove that $W_\textrm{RA}^{*} - W_\textrm{SF4}^{*}>0$. Obviously, this inequality will also hold if $\tau_A>\hat{\tau}_{A,mix}$.

From \cite{Erland/Kaupuzs/Frette/Pugatch/Bachmat:2019} we have that $W_\textrm{SF}^*$ can be represented by one of four different subfunctions,
\begin{widetext} 
	\begin{align}\label{eq:W*_SF}
	\begin{array}{ll}
	W_\textrm{SF1}^* = \frac{\tau_S}{\sqrt{k}} \left[kp(1-C) + kC +1 + C\ln\left(\frac{C}{1+C}\right)  - \ln\left(\frac{2}{1+C}\right) \right]
	&\max\left\{C_2,C_1\right\} \leqslant C \\[0.5ex]
	W_\textrm{SF2}^* = \frac{\tau_S}{\sqrt{k}} \left[kp +1  - \ln\left(\frac{2}{1+C^2(e^{k(1-p)}-1)}\right) \right]
	&C_3^2 \leqslant C^2 \leqslant C_1^2\\[0ex]
	W_\textrm{SF3}^* = \frac{\tau_S}{\sqrt{k}}C\left[k+1-\ln(2)  + \frac{\sqrt{(e^{kp}-1)(1-C^2)}}{C} -\ln\left(1+\frac{\sqrt{(e^{kp}-1)(1-C^2)}}{C}\right)\right]
	&C_4^2 \leqslant C^2 \leqslant C_2^2 \\[0.5ex]
	W_\textrm{SF4}^* = \frac{\tau_S}{\sqrt{k}} \sqrt{(e^{kp}-1) + C^2(e^k - e^{kp})}
	&C^2 \leqslant \min\{C_3^2,C_4^2 \}. 
	\end{array}
	\end{align}
\end{widetext}
Here $C\equiv \tau_F/\tau_S \in(0,1)$, and the borders between the different subdomains as shown in \cref{fig:SF_activesubfunctions} are given by
\begin{align}\label{eq:SFboundaries}
	\begin{array}{lll}
		\textrm{SF1-SF2:} &C_1 	\equiv (e^{k(1-p)}-1)^{-1}\\
		\textrm{SF1-SF3:} &C_2 	\equiv 2e^{-kp}-1\\
		\textrm{SF2-SF4:} &C_3^2 	\equiv (2-e^{kp})/(e^k-e^{kp})\\
		\textrm{SF3-SF4:} &C_4^2	\equiv 4(e^{kp}-1)/(e^{2k} - 4(e^k-e^{kp})).
	\end{array}
\end{align}
For $k<\ln(2)$, both $C_3>1, C_4>1$, and then $W_\textrm{SF}^*=W_\textrm{SF4}^*$. For $k\geqslant\ln(2)$ all subfunctions are present. The borders meet in a vortex point where $p=p^*\equiv \ln[2/(1+C^*)]/k$ and $C=C^*\equiv 2e^{-k}$.

\subsection{Proof \cref{th:RAvsSF}: $W^*_\textrm{RA}-W^*_\textrm{SF}>0$ when $k\leqslant\ln(2)$}\label{proof:RSF4_ksmall}
We show that the difference of the squares is decreasing in $C$ for all $p,k$. It is straightforward to show that the difference is 0 both for $C=1$, $p=0$ and $p=1$. Since the difference is 0 for $C=1$, the difference is positive for $C<1$. 
\begin{align}
\frac{k}{\tau_S^2}(W_\textrm{RA}^{*2} - W_\textrm{SF4}^{*2})
&= \left[p + C^2(1-p)\right](e^{k}-1) -  (e^{kp}-1) - C^2(e^k-e^{kp})\label{eq:RAvsSF4kleq0}\\[0ex]
\frac{k}{\tau_S^2}\frac{\partial (W_\textrm{RA}^{*2} - W_\textrm{SF4}^{*2})}{\partial (C^2)}
&= (1-p)(e^{k}-1) - (e^k-e^{kp}) \equiv g(k,p)\nonumber\\[0ex]
\frac{\partial g}{\partial p}
&= -(e^{k}-1) + ke^{kp}\nonumber\\[0ex]
\frac{\partial ^2 g}{\partial p^2}
&= k^2e^{kp} \;>\;0.\nonumber
\end{align}
Since $g(k,0)=0=g(k,1)$, $g(k,p)<0$ for all $p\in(0,1)$. There is also negative curvature in the $p$-direction:
\begin{align*}
\frac{k}{\tau_S^2}\frac{\partial (W_\textrm{RA}^{*2} - W_\textrm{SF4}^{*2})}{\partial p}
&= (1-C^2)(e^{k}-1-ke^{kp})\\[0ex]
\frac{k}{\tau_S^2}\frac{\partial^2 (W_\textrm{RA}^{*2} - W_\textrm{SF4}^{*2})}{\partial p^2}
&= -(1-C^2)k^2e^{kp} \;<\;0.\\[0ex]
\end{align*}

\subsection{Proof \cref{th:RAvsSF}: necessity of concave mixture}
The necessity of concavity of the aisle-clearing time mixture is shown by a counter-example for small $k$.  In \cref{eq:RAvsSF4kleq0}, it is assumed that $\tau_A^2= \hat{\tau}^2_{A,mix}\equiv p\tau_S^2+(1-p)\tau_F^2$. Instead, assume that the mixture is non-concave, i.e., for $\varepsilon>0$, let $\tau_A^2 \;=\; (1-\varepsilon) \hat{\tau}^2_{A,mix} <  \hat{\tau}^2_{A,mix}$. This gives
\begin{align}
\frac{1}{\tau_S^2}(W_\textrm{RA}^{*2} - W_\textrm{SF4}^{*2})
&= -\varepsilon \hat{\tau}^2_{A,mix} + O(k),
\end{align}
which means that the boarding time will be less for random boarding than for slow first for sufficiently small $k$.

\subsection{Proof \cref{th:RAvsSF}: $W^*_\textrm{RA}-W^*_\textrm{SF}>0$ when $k>\ln(2)$ }\label{ass:proof_kpgtln1}
It is straightforward to show that the difference is 0 both for $C=1$, $p=0$ and $p=1$.

\subsubsection{RA vs. SF1.}
The difference has negative curvature in $p$,
\begin{align*}
\frac{\sqrt{k}}{\tau_S}(W_\textrm{RA}^*- W_\textrm{SF1}^*)
&= \sqrt{C^2 + p(1-C^2)}\left[k+1-\ln(2)\right] - k\left[C + p(1-C)\right]\\[0ex]
&\qquad  - 1 + \ln(2) - C\ln(C)  - (1-C)\ln(1+C)\\[0ex]
\frac{\sqrt{k}}{\tau_S}\frac{\partial (W_\textrm{RA}^*- W_\textrm{SF1}^*)}{\partial p}
&= \frac{(1-C^2)\left[k+1-\ln(2)\right]}{2\sqrt{C^2 + p(1-C^2)}} - k(1-C)\\[0ex]   
\frac{\sqrt{k}}{\tau_S}\frac{\partial^2 (W_\textrm{RA}^*- W_\textrm{SF1}^*)}{\partial p^2}
&= -\frac{(1-C^2)^2\left[k+1-\ln(2)\right]}{4\left[C^2 + p(1-C^2)\right]^\frac{3}{2}} \;<\;0.
\end{align*}
That the difference is positive depends on non-negative difference on the SF3-SF1 and SF1-SF2 border, respectively (see \cref{sssec:RvsSF2,sssec:RvsSF3} below).

\subsubsection{RA vs. SF2.}\label{sssec:RvsSF2}
The difference has negative curvature in $p$,
\begin{align*}
\frac{\sqrt{k}}{\tau_S}(W_\textrm{RA}^*- W_\textrm{SF2}^*)
&= \sqrt{C^2 + p(1-C^2)}\left[k+1-\ln(2)\right] - kp -1 + \ln(2) \\[0ex]
&\qquad  -\ln\left[1+C^2(e^{k(1-p)}-1)\right]\\[0ex]
\frac{\sqrt{k}}{\tau_S}\frac{\partial (W_\textrm{RA}^*- W_\textrm{SF4}^*)}{\partial p}
&= \frac{(1-C^2)\left[k+1-\ln(2)\right]}{2\left[C^2 + (1-C^2)p\right]^\frac{1}{2}} - k
+ \frac{kC^2e^k}{C^2e^k + (1-C^2)e^{kp}}\\[0ex]   
\frac{\sqrt{k}}{\tau_S}\frac{\partial^2 (W_\textrm{RA}^*- W_\textrm{SF4}^*)}{\partial p^2}
&= -\frac{(1-C^2)^2\left[k+1-\ln(2)\right]}{4\left[C^2 + (1-C^2)p\right]^\frac{3}{2}}
- \frac{k^2 C^2 e^k (1-C^2)e^{kp}}{\left[C^2e^k + (1-C^2)e^{kp}\right]^2} \;<\;0.
\end{align*}
That the difference is positive depends on non-negative difference on the SF1-SF2 (smooth) and SF4-SF2 border (see \cref{sssec:RvsSF3}) and that the difference is zero for $p=1$.

\subsubsection{RA vs. SF3.}\label{sssec:RvsSF3}
Set $y\equiv e^{kp}-1$. Then $y\in (0,1)$, since $kp<\ln(2)$. Set $R\equiv \sqrt{(e^{kp}-1)(1-C^2)}/C=\sqrt{y(1-C^2)}/C>0$. To stay in SF3 towards the SF3-SF1 border, $y\leqslant R/(2+R)$, and towards the SF3-SF4 border $R\leqslant (e^k-2)/2$. This means that $R$ is constant at the SF3-SF4 border for fixed $k$.

The re-parameterization gives that
\begin{align}
\frac{\sqrt{k}}{C\tau_S}(W_\textrm{RA}^*- W_\textrm{SF3}^*)
&=(k+L)\left(\sqrt{1+\frac{R^2\ln(1+y)}{ky}} -1\right) +\ln(1+R) -R \nonumber\\
&\equiv f(y,R,k), \label{eq:RAvsSF3}
\end{align}
where $L\equiv 1-\ln(2)$. 

We first find the values of $k\in(\ln(2),\infty)$ that minimizes $f$ in \cref{eq:RAvsSF3} for fixed values of $A\equiv R^2\ln(1+y)/y$. 
\begin{align*}
\frac{\partial f}{\partial k}
& = \frac{A(A-4L)
	\left(k-\frac{L}{1-\sqrt{\frac{4L}{A}}}\right)
	\left[\sqrt{A}(k-L)+2k\sqrt{L}\right]
	}
	{4k^4(\sqrt{A}+2\sqrt{L})
	\sqrt{1+\frac{A}{k}}
	\left[1+\frac{A}{k}\left(1-\frac{k+L}{2k}\right) + \sqrt{1+\frac{A}{k}}\right]}.
\end{align*}
Lower bounds for $f$ on the SF3 domain is given by three different values of $k$, depending on the value of $A=A(R,y)$: 
\begin{itemize}
	\item When $A\leqslant 4L$, $\frac{\partial f}{\partial k}<0$, and $f$ is minimized when $k\rightarrow k_1 \equiv\infty$.
	\item When $A>4L$, $f$ is minimized when $\frac{\partial f}{\partial k}=0$, which is obtained when $k$ is set to 
	\begin{align*}
	k_2(A) \equiv \frac{L}{1-\sqrt{\frac{4L}{A}}}.
	\end{align*}
	Notice that $k_2(A)$ decreases towards $L$ when $A$ increases.
	\item When $A>4L/(1-L/\ln(2))^2\approx 3.95$, then $k_2<\ln(2)$, and $k=k_3\equiv\ln(2)$ gives a lower bound for $f$ since $k>\ln(2)$ by default and $\frac{\partial f}{\partial k}>0$ when $k>k_2$. 
\end{itemize}

Since $\ln(1+y)/y$ is decreasing in $y$, $\partial f/\partial y<0$. Further, since $y\leqslant R/(R+2)$ on the SF3-domain, lower bounds for $f$ can be found by setting $y=y_1\equiv R$ (used below for small $R$) or $y=y_2\equiv 1$ (used below for large $R$) since both $y_1$ and $y_2$ are larger than $R/(R+2)$. Given the fixed value of $R$, $A(R,y(R))$ determines which of the three values of $k$ above that should be chosen to obtain a lower bound for $f$: 

\begin{itemize}
	\item Assume $A\leqslant 4L$, and set $k=k_1=\infty$ and $y=y_1=R$. 
Then $G_1(R)\equiv f(y=R,R,k=\infty)$ is a lower bound for $f$ on SF3 when $R<1.40$, since then $A=R\ln(1+R)<4L$, and 
\begin{align*}
G_1(R) &=\frac{A}{2} + \ln(1+R) - R\\ 
&=\frac{R}{2}\ln(1+R) + \ln(1+R) - R\\
G''_1(R) &= \frac{1}{2(1+R)^2}>0.
\end{align*} 
Since $G_1(0)=G'_1(0)=0$, then $G_1(R)>0$ when $R>0$, and in particular for $R\in (0,1.40)$.

\item Assume $A>4L$, and set $k=k_2$ and $y=y_2=1$. Then $A=R^2\ln(2)>4L$ when $R>2\sqrt{L/\ln(2)}\approx 1.33$. 
Also, with this choice of $y$, $k_2=LR/[R-2\sqrt{L/\ln(2)}]>\ln(2)$ when $R<2\sqrt{L\ln(2)}/(2\ln(2)-1)\approx 2.38$. 
Then $G_2(R)\equiv f(y=1,R,k_2)$ is a lower bound for $f$ on SF3, at least when $R> 1.35$, and 
\begin{align*}
G_2(R) &= 2L\left(\sqrt{\frac{A}{L}}-1\right) + \ln(1+R) - R\\ 
&=2L\left(R\sqrt{\frac{\ln(2)}{L}}-1\right) + \ln(1+R) - R\\
G''_2(R) &= -\frac{1}{(1+R)^2}<0.
\end{align*} 
Since $G_2(1)\approx 0.0018, G_2(3)\approx 0.54$, then $G_2(R)>0$ when $R\in [1,3]$, 
and in particular for $R\in [1.35, 3]$.

\item Assume $A>4L/(1-L/\ln(2))^2$ (which is the case when $y=y_2=1$ and $R>2.38$), 
and set $k=k_3=\ln(2)$. 
Then $G_3(R)\equiv f(y=1,R,k_3)$ is a lower bound for $f$ on SF3 when $R>2.38$, and
\begin{align*}
G_3(R) &= \left(\sqrt{1+\frac{A}{\ln(2)}}-1\right) + \ln(1+R) - R\\ 
&=\left(\sqrt{R^2 + 1}-1\right) + \ln(1+R) - R\\
G'_3(R) &= \frac{R}{\sqrt{1+R^2}}  - \frac{R}{{1+R}} >0,
\end{align*} 
and since $G_3(0)=0$, we get that $G_3(R)>0$ when $R>0$, and in particular for $R>2.38$.
\end{itemize}

\subsubsection{RA vs. SF4.}
We show that the difference of the squares is decreasing in $C$ for all $p,k$,
\begin{align}
\frac{k}{\tau_S^2}(W_\textrm{RA}^{*2} - W_\textrm{SF4}^{*2})
&= \left[p + C^2(1-p)\right][k-\ln(2)+1]^2 -  (e^{kp}-1) - C^2(e^k-e^{kp})
\label{eq:SF4squareddiff}\\
&\equiv f(k,p,C)\nonumber\\[0ex]
\frac{k}{\tau_S^2}\frac{\partial (W_\textrm{RA}^{*2} - W_\textrm{SF4}^{*2})}{\partial (C^2)}
&= (1-p)[k-\ln(2)+1]^2 - (e^k-e^{kp}) \nonumber\\[0ex] 
&\leqslant (1-p)(e^{k}-1) - (e^k-e^{kp}) \equiv g(k,p) \;<\;0. \label{eq:SF4partialC2}
\end{align}
The first inequality in \cref{eq:SF4partialC2} is due to
\begin{align*}
h(k) &\equiv (e^k-1)  - [k-\ln(2)+1]^2\\[0ex]
h'(k) &= e^k  - 2[k-\ln(2)+1]\\[0ex]
h''(k) &= e^k  - 2 > 0
\end{align*}
when $k>\ln(2)$. Since $h(\ln(2))=0=h'(\ln(2))$, $h(k)>0$ when $k>\ln(2)$. The second inequality in \cref{eq:SF4partialC2} was shown in \cref{proof:RSF4_ksmall} for all $k>0, p\in(0,1)$. 

There is also negative curvature in the $p$-direction,
\begin{align*}
\frac{k}{\tau_S^2}\frac{\partial (W_\textrm{RA}^{*2} - W_\textrm{SF4}^{*2})}{\partial p}
&= (1-C^2)\lbrace[k+1-\ln(2)]^2-ke^{kp}\rbrace\\[0ex]
\frac{k}{\tau_S^2}\frac{\partial^2 (W_\textrm{RA}^{*2} - W_\textrm{SF4}^{*2})}{\partial p^2}
&= -(1-C^2)k^2e^{kp} \;<\;0.
\end{align*}
Due to the negative curvature in $p$, the difference in \cref{eq:SF4squareddiff} is positive on SF4 if it is positive on the SF3-SF4 border (proved in \cref{sssec:RvsSF3}) and on the SF4-SF2 border. 

At the latter border $p<\ln(2)/k$, and $C<C^*\equiv 2e^{-k}$. Due to the negative curvature in $p$, it is sufficient to show that the difference is positive on a line extending from $C=0$ to $C=C^*$ with fixed $p=\ln(2)/k$. Since the difference is decreasing in $C$ (\cref{eq:SF4partialC2}), it is sufficient to show that the difference is positive at the point $(p=\ln(2)/k, C=C^*)$,
\begin{align*}
f\left(k,p=\frac{\ln(2)}{k},C=C^*\right)
&=\frac{1}{k}\left[l_2+4e^{-2k}(k-l_2)\right]\left[k-l_2+1\right]^2 - 1- 4e^{-2k}(e^k-2)\equiv F(k),
\end{align*}
where $l_2\equiv \ln(2)$. Let $K\equiv k-l_2$ (such that $k>\ln(2)$ implies $K>0$), then
\begin{align*}
\frac{ke^{2K} F(k)}{K}
&= e^{2K}(Kl_2+2l_2-1) +(K+1)^2 -2(e^K-1) - 2l_2\frac{e^{K}-1}{K} \equiv h(K).
\end{align*}
Since $h(0)=0$, it is sufficient to show that $h'(K)>0$ to prove that $h(K)> 0$ for $K>0$ (which in turn implies that $F(k)> 0$ for $k>\ln(2)$),
\begin{align}\label{eq:RvsSF4_hdiff}
\frac{h'(K)}{2}
&= \underbrace{e^{2K} (Kl_2+2.5l_2-1)  + (K+1)  -e^{K}}_{g_1(K)} 
- \underbrace{{l_2}\frac{e^{K}(K-1) + 1}{K^2}}_{g_2(K)}.
\end{align}
For $K\leqslant 1$, the first term $g_1(K)$ is increasing. Hence, a lower bound is $g_1(K)\geqslant g_1(0)=0.73$ when $K\geqslant 0$.
An upper bound for $g_2(K)$ is found by
$e^K \geqslant 1 + K + K^2/2$ and that the factor $(K-1)$ is negative for $0<K<1$. Then we can write
for $0<K<1$
\begin{align*}
g_2(K) \leqslant l_2 \frac{(1+K+K^2/2)(K-1) + 1}{K^2} 
= \frac{l_2}{2}(1 + K) \equiv \tilde{g}_2(K).
\end{align*}
Using that this upper bound is increasing, 
$g_2(K)\leqslant\tilde{g}_2(1)=l_2=0.69 $ when $K\leqslant 1$.
This implies that $g_1(K)-g_2(K)>0.73-0.69>0$ for $K\leqslant 1$. For $K>1$, it is straightforward to show that each of the negative terms in \cref{eq:RvsSF4_hdiff} are dominated by one of the respective positive terms.

\subsection{Proof of \cref{eq:RAvsFFmaxmain}: maximum relative distance between RA and SF policies}
We propose that the maximum relative distance $(W^*_\textrm{RA}-W^*_\textrm{SF})/W^*_\textrm{SF}=W^*_\textrm{RA}/W^*_\textrm{SF}-1$ is in the SF4-region.
For fixed $k>\ln(2)$, we therefore seek the maximum of
\begin{align}\label{eq:RArelSF4squared}
\frac{1}{\left[k-\ln(2)+1\right]^2}\left(\frac{W^*_\textrm{RA}}{W^*_\textrm{SF4}}\right)^2
&=\frac{p+C^2(1-p)}{e^{kp}-1 + C^2(e^k - e^{kp})}
=\frac{C^2+p(1-C^2)}{e^{kp}(1-C^2) + C^2 e^k - 1}\equiv g(k,p,C).
\end{align}
The same equation applies for $k\leqslant\ln(2)$, except that the leftmost denominator is exchanged with $e^k-1$.

Partial differentiation of $g$ gives
\begin{align}\label{eq:RArelSF4squaredpartialC2}
\frac{\partial g}{\partial (C^2)}
&=\frac{e^{kp}-1 -p(e^k-1)}{\left[e^{kp}-1 + C^2(e^k - e^{kp})\right]^2}  \leqslant 0,
\end{align}
since the numerator of ${\partial g}/{\partial (C^2)}$ has positive curvature in $p$ and equals 0 both for $p=0$ and $p=1$. Consequently, for fixed $k,p$ and $C\in(0,1)$, $g$ is maximized when $C\rightarrow 0$.

Partial differentiation of $g$ with respect to $p$ gives
\begin{align}
\frac{1}{1-C^2}\frac{\partial g}{\partial p}
&=\frac{e^{kp}(1-kp)-1  +  C^2\left[e^k - e^{kp}(k-kp+1)\right]}{\left[e^{kp}-1 + C^2(e^k - e^{kp})\right]^2}. \label{eq:partial_gp}
\end{align}
A local optimum if found by setting \cref{eq:partial_gp} to zero. This gives
\begin{align}
C^2
&=\frac{1-e^{kp}(1-kp)}{e^k - e^{kp}\left[k(1-p)+1\right]}\label{eq:Coptimum}\\
&\overset{p\rightarrow 0}{\sim} \quad \frac{(kp)^2}{e^k-k+1}
\quad \overset{k\rightarrow \infty}{\sim} \quad \frac{(kp)^2}{e^k}. \label{eq:Coptimumasympt}
\end{align}
For fixed $k$, the numerator in \cref{eq:Coptimum} is increasing from 0 when $p>0$, and the denominator is non-negative and bounded from above. Hence, $p$ must be small to minimize $C$.

The asymptotic optimal value for $C^2$ in \cref{eq:Coptimumasympt} can be inserted into \cref{eq:RArelSF4squared}, which gives the maximum
\begin{align}
\left(\frac{W^*_\textrm{RA}}{W^*_\textrm{SF4}}\right)^2
&=\left[k-\ln(2)+1\right]^2\frac{p+\frac{(kp)^2}{e^k-k+1}(1-p)}{e^{kp}-1 + \frac{(kp)^2}{e^k-k+1}(e^k - e^{kp})}\nonumber\\
&\overset{p\rightarrow 0}{\sim} \quad
\frac{\left[k-\ln(2)+1\right]^2}{k} \frac{(e^k-k+1) +k^2p}{(e^k-k+1) +kp(e^k-1)} \nonumber\\
&\overset{k\rightarrow \infty}{\sim} 
\quad k.\nonumber
\end{align}
This means that for fixed $k$, the maximal relative difference between random boarding and slow first equals $\left[k-\ln(2)+1\right]/\sqrt{k}-1$ and is obtained for  $p$ small and $C\approx kp/\sqrt{e^k-k+1}$. 

For large $k$, the maximal relative difference between random boarding and slow first is obtained for $p$ small and $C\approx kp/\sqrt{e^k}$, and by increasing $k$ the relative difference can be infinitely large,
\begin{align*}
\frac{W^*_\textrm{RA}-W^*_\textrm{SF4}}{W^*_\textrm{SF4}}
& \approx \sqrt{k}-1.
\end{align*}
{Theoretically, there could be other local maxima in other subdomains. However, numerical inspections indicate that the given solution is the global maximum.}

\subsection{Proof of \cref{th:taubounds}: bounds on the effective aisle-clearing time $\tau_X$}\label{apps:taubounds}
Since $\tau_X$ is independent of $k$, we look at the case $k=0$ where the weight of the heaviest chain is asymptotically $2\tau_X \sqrt{N}$ for large $N$. We apply the Vershik-Kerov theorem which states that when $X$ is deterministic and $k=0$, the number of points in a longest chain is asymptotically $2\sqrt{N}$. 

For $\tau_X$ and a given $u\in [a,b]$, the lower bound is established by considering the weight of the longest sequence of points 
with weight at least $u$. There are roughly $\int_u^b f(t)dt N$ such points and so by the Vershik-Kerov theorem we have a longest chain of size roughly $2\sqrt{\int_u^bf(t)dt}\sqrt{N}$ points. The average weight of a point in this chain is the same as the average weight of a point conditioned on being larger than $u$, i.e. $\int_u^btf(t)dt/\int_u^bf(t)dt$. Since this holds for any $u$, the result follows.

The upper bound on $\tau_X$ is obtained by replacing $X$ by $X_u$ that takes the values $u_i$ with probability $p_i=Pr(u_{i-1}< X\leq u_i)$. Obviously $\tau_X\leq \tau_{X_u}$, since $X_u$ dominates $X$. Consider the heaviest chain with respect to $X_u$. The number of points with weight $u_i$ in this chain is bounded from above by the size of longest chain of such points which by Vershik-Kerov has a total weight of roughly $2 u_i\sqrt{p_i}\sqrt{N}$. Adding this up gives that the weight of the heaviest chain is estimated from above by $2(\sum_{i=1}^n u_i\sqrt{p_i})\sqrt{N}$.  
The upper bound in \cref{eq:taubounds} is obtained since this holds for all subdivisions. 

For the lower bound of $\sqrt{\langle X^2 \rangle}$, we apply the Cauchy-Schwartz inequality $\int_a^bg(t)f(t)dt\leq \sqrt{\int_a^bg^2(t)dt} \sqrt{\int_a^bh^2(t)dt}$. For a given $u\in [a,b]$, let $g=1_{[u,b]}\sqrt{f(t)}$ and $h=t\sqrt{f(t)}$ (here $1_{[u,b]}$ is the indicator of the interval $[u,b]$ which equals $1$ for points in the interval and $0$ otherwise). For these $f$ and $g$ the two sides of the lower bound inequality coincide with the two sides of Cauchy-Schwartz.

Regarding the upper bound of $\sqrt{\langle X^2 \rangle}$, we note that since the $\ell_1$ norm of a vector is greater or equal to its $\ell_2$ norm
\begin{equation*}
\sum_{i=1}^n u_i\sqrt{p_i}
\geqslant \sqrt{\sum_{i=1}^n u_i^2p_i}
=\sqrt{\langle X_u^2 \rangle} 
\geqslant \sqrt{\langle X^2 \rangle}.
\end{equation*}

\subsection{Asymptotic relative distance between RA and SF policies for $k\rightarrow\infty$}\label{app:RAvsSF_kinfty}
For large $k$, the SF1 domain dominates the $p,C$ unit square which follows straightforwardly from the SF1-boundaries in \cref{eq:SFboundaries}. For fixed $p,C$, the relative distance is independent of $k$ for large $k$ and \cref{eq:RArelSFkinfty} follows since
\begin{align}\label{eq:RArelSF1}
\frac{W^*_\textrm{RA}}{W^*_\textrm{SF1}}
&=\frac{\sqrt{p+C^2(1-p)}\left[k+1-\ln(2)\right]}{\left[p+C^2(1-p)\right]k + 1 - \ln(2) + (1-C)\ln(1+C) + C\ln(C)}\\
&\overset{k\rightarrow \infty}{\longrightarrow} \quad 
\frac{\sqrt{p+C^2(1-p)}}{p+C(1-p)} 
\equiv h(p,C) \label{eq:RArelSF1_h}\\
&= 1+ \frac{p(1-p)\left(1-C^2\right)}{\left[p+C(1-p)\right]\left[p+C(1-p) + \sqrt{p+C^2(1-p)}\right]} >1.\nonumber
\end{align}
The last equality comes from the identity $a/b\equiv 1+(a^2-b^2)/[b(a+b)]$.
The partial derivatives of $h$ are,
\begin{align}
\frac{\partial h}{\partial C}
&=-\frac{p(1-p)(1-C)}{2\sqrt{p+C^2(1-p)}\left[p+C(1-p)\right]^2}\;<\;0 \label{eq:RArelSF1partialC}\\
\frac{\partial h}{\partial p}
&=\frac{(1-C)^2\left[C-p(1-C)\right]}{2\sqrt{p+C^2(1-p)}\left[p+C(1-p)\right]^2}.\nonumber
\end{align}
This gives that the asymptotic function $h(p,C)$ in \cref{eq:RArelSF1_h} is increasing with decreasing $C$. For fixed $C$, the maximum is at $p=C/(1-C)$ with maximum value $h^*=(1+C)/(2\sqrt{C})$.

For fixed $p,C$, the partial derivative with respect to $k$ gives that the relative difference in \cref{eq:RArelSF1} is decreasing with increasing $k$ towards the asymptotic function $h(p,C)$, if
\begin{align}\label{eq:RvsSF1decrease}
p &> 1 + \frac{1}{1-\ln(2)}\left[\ln(1+C) + \frac{C\ln(C)}{1-C}\right].
\end{align}
This means that $h(p,C)$ is a lower bound for ${W^*_\textrm{RA}}/{W^*_\textrm{SF1}}$ in this area of the $p,C$ unit square for all $k>\ln(2)$.
On the other hand, if $p$ is smaller than the right-hand side of \cref{eq:RvsSF1decrease}, the relative difference is increasing towards $h(p,C)$. 


%

\end{document}